\documentclass[aps,superscriptaddress,reprint,nofootinbib]{revtex4-1}

\usepackage{graphicx}
\usepackage{amsmath}
\usepackage{amssymb}
\usepackage{amsfonts}
\usepackage{filecontents}
\usepackage{color}
\usepackage[normalem]{ulem} 
\usepackage{soul}
\usepackage{epstopdf}
\usepackage{nameref}
\usepackage{hyperref}
\hypersetup{hidelinks}
\usepackage{fancyhdr}
\usepackage{float}
\usepackage{gensymb}
\usepackage{physics}
\usepackage{upgreek}
\newcounter{fig}
\usepackage{lipsum}
\usepackage{natbib}
\usepackage{bibunits}
\usepackage{makecell}
\usepackage[usenames,dvipsnames,svgnames,table]{xcolor}
\usepackage{enumitem}
\usepackage{multirow}
\usepackage{mathtools}

\DeclareRobustCommand{\SkipTocEntry}[5]{}

\renewcommand{\bibliography}[1]{} % references are embedded below; suppress RevTeX external .bbl loading

\begin{document}
\begin{bibunit}

\title{
Robust biexciton preparation for entangled photon-pair generation by single-shot Floquet driving
}

\newcommand{\affZJU}{
State Key Laboratory of Extreme Photonics and Instrumentation,
College of Information Science and Electronic Engineering,
Zhejiang University, Hangzhou 310027, China
}

\newcommand{\affBayreuth}{
Theoretische Physik III, Universität Bayreuth,
95440 Bayreuth, Germany
}

\newcommand{\affRUB}{
Experimental Physics VI, Ruhr University Bochum,
Bochum, Germany
}

\newcommand{\affHCU}{
Cryogenic Center, Hangzhou City University,
Hangzhou 310015, China
}

\newcommand{\affCryogenicZJU}{
Zhejiang Key Laboratory of Refrigeration and Cryogenic Technology,
Zhejiang University, Hangzhou 310027, China
}

\newcommand{\affZJUGlobal}{
ZJU-Hangzhou Global Scientific and Technological Innovation Center,
Zhejiang University, Hangzhou, China
}

\newcommand{\affMicroNano}{
Zhejiang Key Laboratory of Micro-Nano Quantum Chips and Quantum Control,
School of Physics, Zhejiang University, Hangzhou 310027, China
}

\newcommand{\affDortmund}{
Condensed Matter Theory, Department of Physics,
TU Dortmund, 44227 Dortmund, Germany
}

\newcommand{\affPresent}{
Present address: Department of Electrical and Computer Engineering,
National University of Singapore, Singapore, Singapore
}

\affiliation{\affZJU}         % 1
\affiliation{\affBayreuth}    % 2
\affiliation{\affRUB}         % 3
\affiliation{\affHCU}         % 4
\affiliation{\affCryogenicZJU}% 5
\affiliation{\affZJUGlobal}   % 6
\affiliation{\affMicroNano}   % 7
\affiliation{\affDortmund}    % 8
\affiliation{\affPresent}     % 9

\author{Jun-Yong Yan}
\thanks{These authors contributed equally to this work.}
\affiliation{\affZJU}
\affiliation{\affPresent}

\author{Paul C. A. Hagen}
\thanks{These authors contributed equally to this work.}
\affiliation{\affBayreuth}

\author{Hang-Yu Ge}
\author{Fang-Yuan Li}
\affiliation{\affZJU}

\author{Hans-Georg Babin}
\affiliation{\affRUB}

\author{\\Wei E. I. Sha}
\affiliation{\affZJU}

\author{Bo Wang}
\affiliation{\affHCU}

\author{Zhi-hua Gan}
\affiliation{\affCryogenicZJU}

\author{Andreas D. Wieck}
\author{Arne Ludwig}
\affiliation{\affRUB}

\author{Chao-Yuan Jin}
\affiliation{\affZJUGlobal}

\author{\\Vollrath M. Axt}
\affiliation{\affBayreuth}

\author{Da-Wei Wang}
\affiliation{\affMicroNano}

\author{Moritz Cygorek}
\email[Email to: ]{moritz.cygorek@tu-dortmund.de}
\affiliation{\affDortmund}

\author{Feng Liu}
\email[Email to: ]{feng\_liu@zju.edu.cn}
\affiliation{\affZJU}

\begin{abstract}

{Quantum emitters driven by resonant two-photon excitation are a leading source for deterministically generated entangled photon pairs, essential for scalable photonic quantum technologies. However, conventional excitation schemes pose various constraints, limiting their scalability and practicability. Here, we demonstrate how biexciton preparation schemes with significantly improved robustness and reduced experimental demands can be identified using a novel design principle: \textit{ultrafast single-shot Floquet driving}. This is achieved by employing two strongly and symmetrically detuned pulses, whose superposition generates a stroboscopic Hamiltonian that enables direct coupling between ground and biexciton states.
Moreover, a pulse delay serves as a tuning knob, introducing an effective magnetic field making biexciton preparation particularly robust. Experimentally, we achieve a biexciton occupation exceeding 96\% and preserve photon-pair entanglement with a fidelity of 93.4\%. Our work opens avenues for the design of quantum control protocols with superior performance across diverse quantum emitters and qubit platforms. 
} 
\end{abstract}

\maketitle
Quantum entanglement, one of the most intriguing phenomena in the quantum world, lies at the heart of quantum information science. A deterministic source of entangled photons is a fundamental building block for modern quantum photonics applications, such as quantum communications~\cite{Zhang2022,Li2019}, optical quantum computing~\cite{Knill2001a,Kimble2008}, quantum imaging and sensing~\cite{Defienne2024,Pirandola2018}. A semiconductor quantum dot (QD), often referred to as an artificial atom due to its three-dimensional electronic confinement and discrete energy levels, is a well-established source of highly entangled photons~\cite{Huber2018b,Chen2024,Chen2018a}. The on-demand high-fidelity preparation of the biexciton state is the starting point for the radiative cascade decay process that generates entangled photon pairs. Moreover, cascade-emitted photons have been shown to induce simultaneous excitation of two independent quantum emitters~\cite{Muthukrishnan2004}, enabling the synthesis of effective multi-qubit interactions and coherent generation of entangled states~\cite{Ren2020}.

\begin{figure}[tbp]
\refstepcounter{fig}
	\includegraphics[width=1\linewidth]{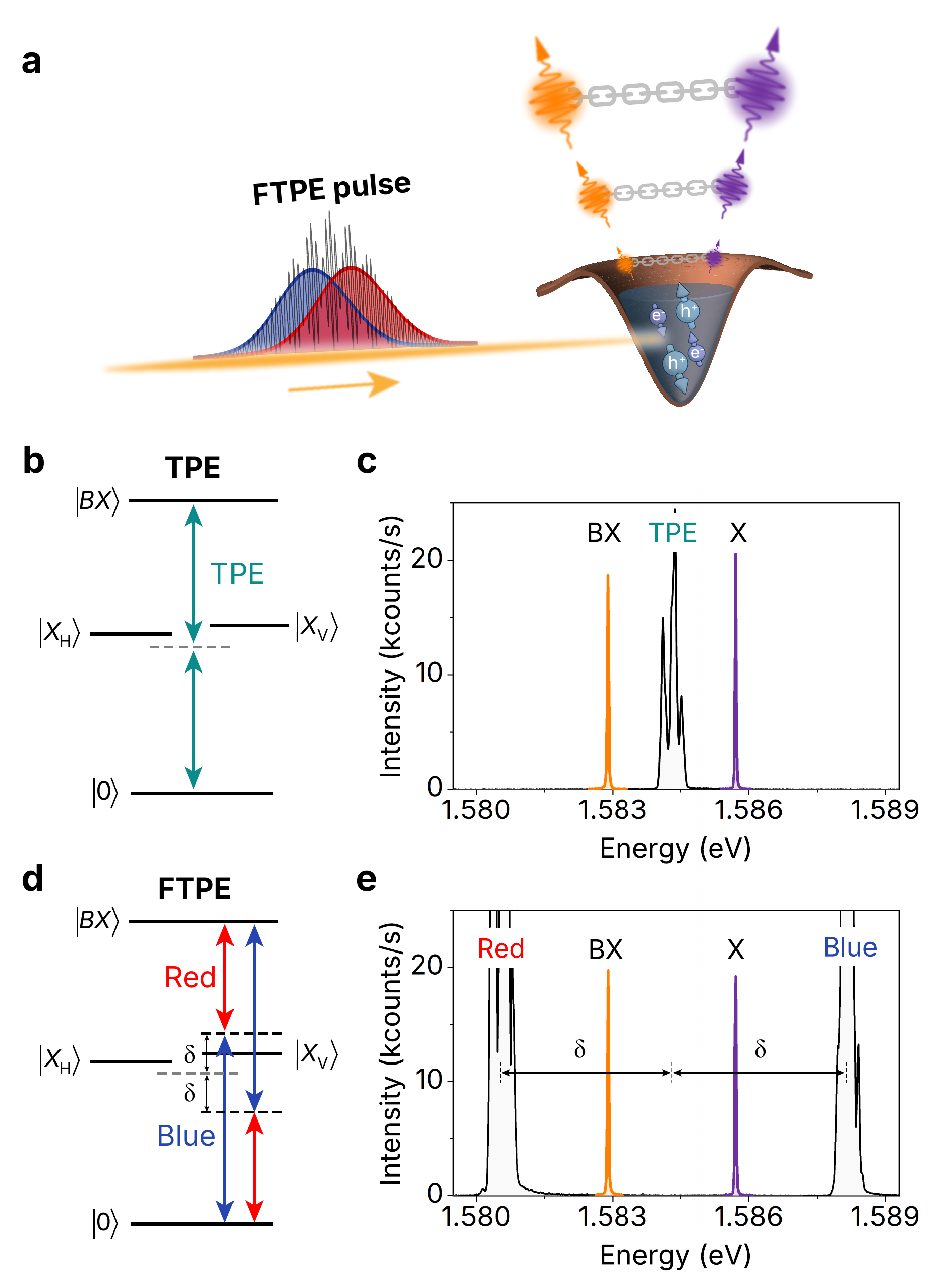}
	\caption{\textbf{TPE and FTPE schemes.}~\textbf{a}, Biexciton state prepared by FTPE emits pairs of entangled photons. \textbf{b}, Schematic of the TPE process. \textbf{c}, Laser excitation spectrum under TPE, where the energy of the TPE pulse is set at the average energy of the \textit{BX} and \textit{X} photons. \textbf{d}, Schematic of FTPE. \textbf{e}, Laser excitation spectrum under FTPE, where the blue and red pulses are detuned symmetrically to meet the two-photon resonance condition. }
	\label{principle}
\end{figure}

A degenerate resonant two-photon excitation (TPE) scheme, where two photons with identical frequencies are absorbed simultaneously, is commonly used to coherently prepare the biexciton state with minimal dephasing and time jitter~\cite{Stufler2006} (see Fig.~\ref{principle}(b)). This excitation scheme has shown significant potential in generating polarization-entangled photon pairs~\cite{Mueller2014,Liu2019e,Wang2019l}, time-bin entangled photons~\cite{Jayakumar2014a,Gines2021} and indistinguishable single photons~\cite{Yan2022S,Sbresny2022,Wei2022}. However, conventional TPE relies on Rabi oscillations of biexciton populations, making its excitation efficiency highly sensitive to variations in the excitation laser power and the dipole moment of the emitters. Furthermore, tuning the laser to the average frequency of the biexciton (\textit{BX}) and exciton (\textit{X}) photons is unfavorable for QDs with small or negative \textit{BX} binding energy. In the former case~\cite{Moroni2019,Sarkar2006}, the tiny detuning between the laser and the entangled photon pairs makes it particularly difficult to perform effective laser filtering. In the latter case, the TPE process competes with phonon-assisted excitation of $X$, reducing the preparation efficiency of $BX$~\cite{Juska2020a}. These limitations hinder the performance and scalability of on-demand entangled photon sources.
Although several alternative approaches based on a single excitation pulse, such as rapid adiabatic passage~\cite{Glassl2013b,Kaldewey2017a,Karli2024} and LA-phonon-assisted excitation~\cite{Quilter2015a,Bounouar2015b,Reindl2017a}, have been explored, the simultaneous realization of robust state preparation and the generation of high-fidelity entangled photon pairs has not yet been achieved. Moreover, while multi-colour excitation schemes such as the swing-up of quantum emitter (SUPER)~\cite{Bracht2023,Heinisch2024,Kuniej2024} for biexciton preparation have been proposed, their experimental implementation for entangled-photon generation has not yet been demonstrated.

Floquet engineering, which employs periodic driving to achieve desired quantum dynamics, offers a fundamentally different route for overcoming the limitations of existing excitation schemes.
According to Floquet's theorem~\cite{Shirley1965}, when a quantum system is periodically driven with a period $T$, the time evolution operator evaluated at stroboscopic times (integer multiples of $T$) takes the same form as for a time-independent Hamiltonian (see Methods Section). 
This stroboscopic Hamiltonian governs the behaviour of the quantum system on a coarse-grained time scale $\gtrsim T$. Moreover, if the driving oscillates periodically within a slowly varying envelope, the system approximately follows the instantaneous stroboscopic Hamiltonian, which is then time-dependent on the scale of the envelope~\cite{Ikeda2022}.  
The ability to control and tune the properties of quantum systems by Floquet engineering has enabled the realization of photonic topological insulators~\cite{Rechtsman2013} and synthetic gauge fields in photonic systems~\cite{Cheng2025} and in ultracold atomic systems~\cite{Weitenberg2021}. 

Existing Floquet engineering approaches typically utilize modulated continuous-wave laser fields or repetitive laser pulse trains. In the context of on-demand semiconductor quantum light sources, Floquet engineering has focused mainly on engineering excitonic wave functions by periodic microwave driving~\cite{Iorsh2022}.
Despite these advances, all-optical Floquet engineering with \textit{single-cycle pulses} to manipulate the optical excitation dynamics remains unexplored. 

In this article, we propose and demonstrate a scheme for robust on-demand biexciton preparation and entangled photon generation enabled by all-optical single-shot Floquet driving [Fig.~\ref{principle}(a)].
Our key observation is that one can realize a time-dependent stroboscopic Hamiltonian that directly drives the ground-to-biexciton transition of a semiconductor quantum dot simply by dichromatic excitation with laser pulses symmetrically detuned from the degenerate two-photon resonance by a detuning $\delta$ [Fig.~\ref{principle}(d) and (e)]. 
Further requirements are that the detuning is large enough so that oscillations with frequency $\delta$ are fast compared to changes in the pulse envelope and that the pulses overlap temporally. Control of the phase relation between the pulses is not required (see Supplementary Sect.~\ref{SI_Sec_phserelation}). These requirements are well achievable within current state-of-the-art experimental techniques for two-colour excitation~\cite{Koong2021,yan2023,Boyle2010a,Yan2024a}. Moreover, the dichromatic pulses can be detuned far from the $X$ and $BX$ photons, relaxing the requirement for laser filtering, especially for QDs with small or negative $BX$ binding energy. 
Hence, our proposed excitation scheme, which we refer to as \textit{Floquet-engineered Two-Photon Excitation} (FTPE), can be readily implemented for a wide range of QDs. In our experiments, we demonstrate a $BX$ preparation efficiency of 96.6\% and a polarization-entanglement fidelity of 93.4\%, proving its reliability.

\begin{figure*}[htbp]
\refstepcounter{fig}
	\includegraphics[width=1\linewidth]{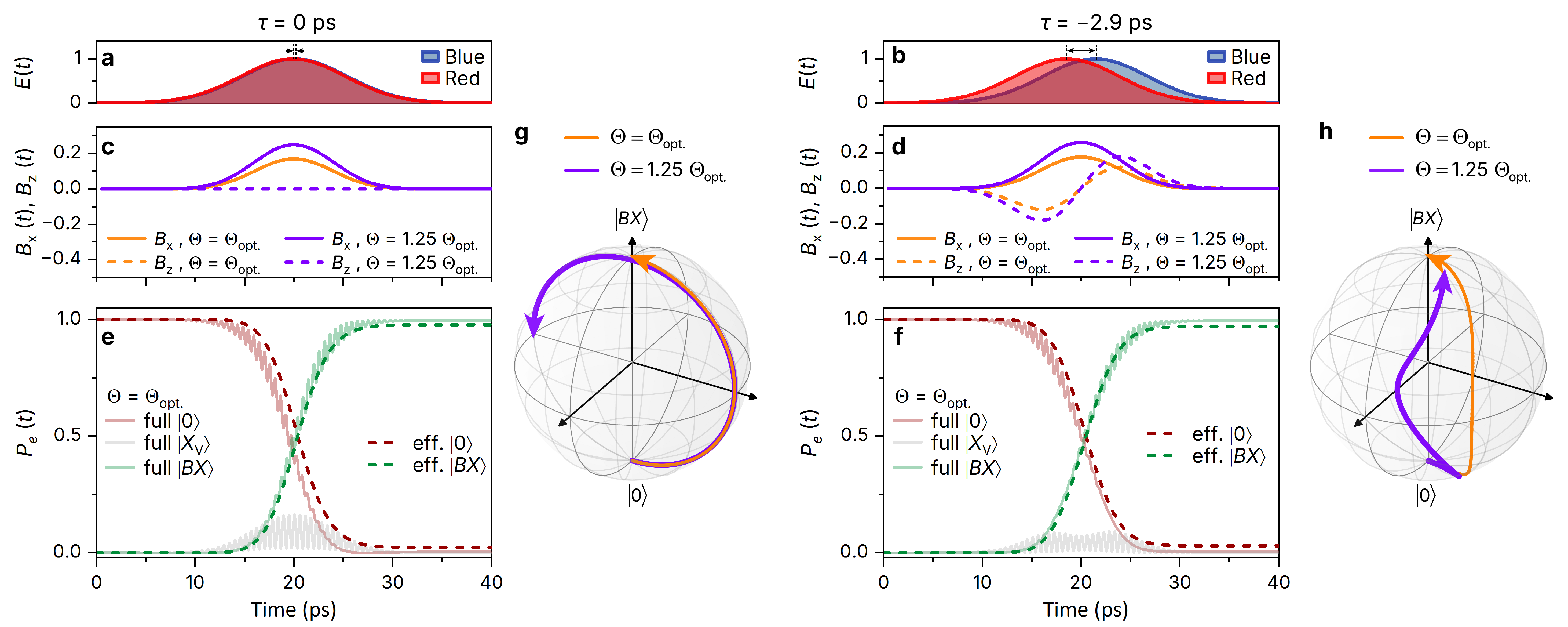}   \caption{\textbf{Simulated FTPE for different pulse delays $\tau$.} \textbf{a} and \textbf{b}, Time-dependent laser amplitudes of red- and blue-detuned pulses for delays $\tau = 0$~ps and $\tau = -2.9$~ps, respectively.  \textbf{c} and \textbf{d}, Time dependence of the effective magnetic field components $B_x(t)$ and $B_z(t)$ in the stroboscopic two-level Hamiltonian in Eq.~(\ref{eq:stroboscopicHamiltonian_mainText}) for delays $\tau=0$ and $-2.9$~ps. For each delay, we show the result obtained at the optimal driving strength $\Theta=\Theta_{\rm opt.}$, together with the result for $\Theta=1.25\Theta_{\rm opt.}$. The detuning $\hbar\delta=3.75$~meV was chosen. 
    \textbf{e} and \textbf{f}, Simulated time evolution of the QD state occupations for the optimal pulse areas corresponding to $\tau=0$ and $-2.9$~ps, respectively. Full: solution from the full Hamiltonian of Eq.~\eqref{eq:originalHamiltonian}. Eff.: solution from the effective stroboscopic model of Eq.~\eqref{eq:stroboscopicHamiltonian_mainText}.~\textbf{g} and \textbf{h}, Corresponding trajectories on the Bloch sphere of the effective two-level system composed of ground and biexciton states.}
    \label{Fig2}
\label{evolution}
\end{figure*}
\addtocontents{toc}{\SkipTocEntry}
\section*{Results}
\addtocontents{toc}{\SkipTocEntry}
\subsection*{Theory on Floquet-engineered two-photon excitation}

Figure~\ref{principle}(a) illustrates the proposed FTPE scheme applied to a semiconductor QD. A pair of temporally partially overlapping pulses (depicted in blue and red) generates a periodic driving field (indicated by gray lines). This periodic driving deterministically and reliably creates a biexciton in the QD, which subsequently emits entangled photon pairs via the cascade decay process.

To model the FTPE process, we consider a ladder system composed of a ground state $\left|0\right>$, a vertically (V) polarized exciton $\left|X_{\rm{V}}\right>$, and a biexciton state $\left|BX\right>$. We assume that both driving lasers are vertically polarized such that the horizontally (H) polarized exciton is not involved in the excitation process. The energy of $\left|BX\right>$ is $E_{BX}=2E_{X}-E_b$, where $E_{X}$ is the energy of the $X$ state and $E_b$ is the biexciton binding energy. Figure~\ref{principle}(b) shows the schematic of conventional TPE, where the laser is tuned to be strictly in resonance with the virtual state of the $\left|0\right>$ to $\left|BX\right>$ degenerate two-photon transition and Rabi oscillations can be observed~\cite{Stufler2006}. A representative TPE laser excitation spectrum is presented in Fig.~\ref{principle}(c). The laser frequency is tuned to the midpoint between the $BX$ and $X$ emission lines. In contrast, as shown in Fig.~\ref{principle}(d), the two colour lasers used in FTPE are detuned to lie outside the $BX$ and $X$ lines. Thus, a large range of laser detunings is allowed as long as the two-photon resonance is maintained~\cite{Varada1992}, i.e., $E_{\mathrm{blue}}$+$E_{\mathrm{red}}$=$E_{BX}$. The higher- and lower-energy lasers are labeled as blue and red pulses, respectively. Since the $\left|BX\right>$ to $\left|X_V\right>$ and $\left|X_V\right>$ to $\left|0\right>$ transitions have the same optical polarization selection rule, each laser pulse simultaneously couples to both transitions [Fig.~\ref{principle}(d)]. Figure~\ref{principle}(e) shows a representative laser excitation spectrum under FTPE. The detuning of either laser from conventional TPE resonance is labeled as $\delta$. Some ripples in the excitation pulse spectra are attributed to the polarization optics used in the setup.
In the frame rotating with the frequency of resonant TPE ($E_{BX}/(2\hbar)$), the Hamiltonian in the basis $\{BX, X_{\rm{V}},0\}$ is given by
\begin{equation} \label{eq:originalHamiltonian}
H/\hbar = \left(\begin{array}{ccc}
    0 & \Omega^*(t)/2 & 0 \\
    \Omega(t)/2 & E_b/(2\hbar) & \Omega^*(t)/2\\
    0 & \Omega(t)/2 & 0
\end{array}\right),
\end{equation}
where the light-matter interaction term contains two equally strong laser pulses $\Omega(t)=f(t-\tau/2)e^{-i\delta (t-\tau/2)}+f(t+\tau/2)e^{i\delta (t+\tau/2)}$, which can be delayed relative to each other by a time $\tau$. The individual pulse envelopes are Gaussian
$f(t)= \frac{\Theta}{\sqrt{2\pi}s}e^{-t^2/2s^2}$
with single-photon transition pulse area $\Theta$ and Gaussian standard deviation $s=\frac{\tau_0}{2\sqrt{2\ln 2}} $ related to the FWHM duration of the electric field $\tau_0 = 12.49$~ps. The pulse area is defined as
$\Theta=\left|\frac{\mu}{\hbar} \int_{-\infty}^{+\infty} \mathcal{E}_0(t) \mathrm{d} t\right|$,
where \(\mu\) is the transition dipole moment for each single-photon transition, and \(\mathcal{E}_0(t)\) is the electric-field amplitude of either the blue- or red-detuned pulse. The full quantum dynamics of the driven QD including phonon effects~\cite{Glassl2013c,Reiter2014a} is simulated using the numerically exact process tensor method \cite{CygorekPRX, Cygorek2024} (see Methods Section). 

\begin{figure*}[htbp]
\refstepcounter{fig}
	\includegraphics[width=0.9\linewidth]{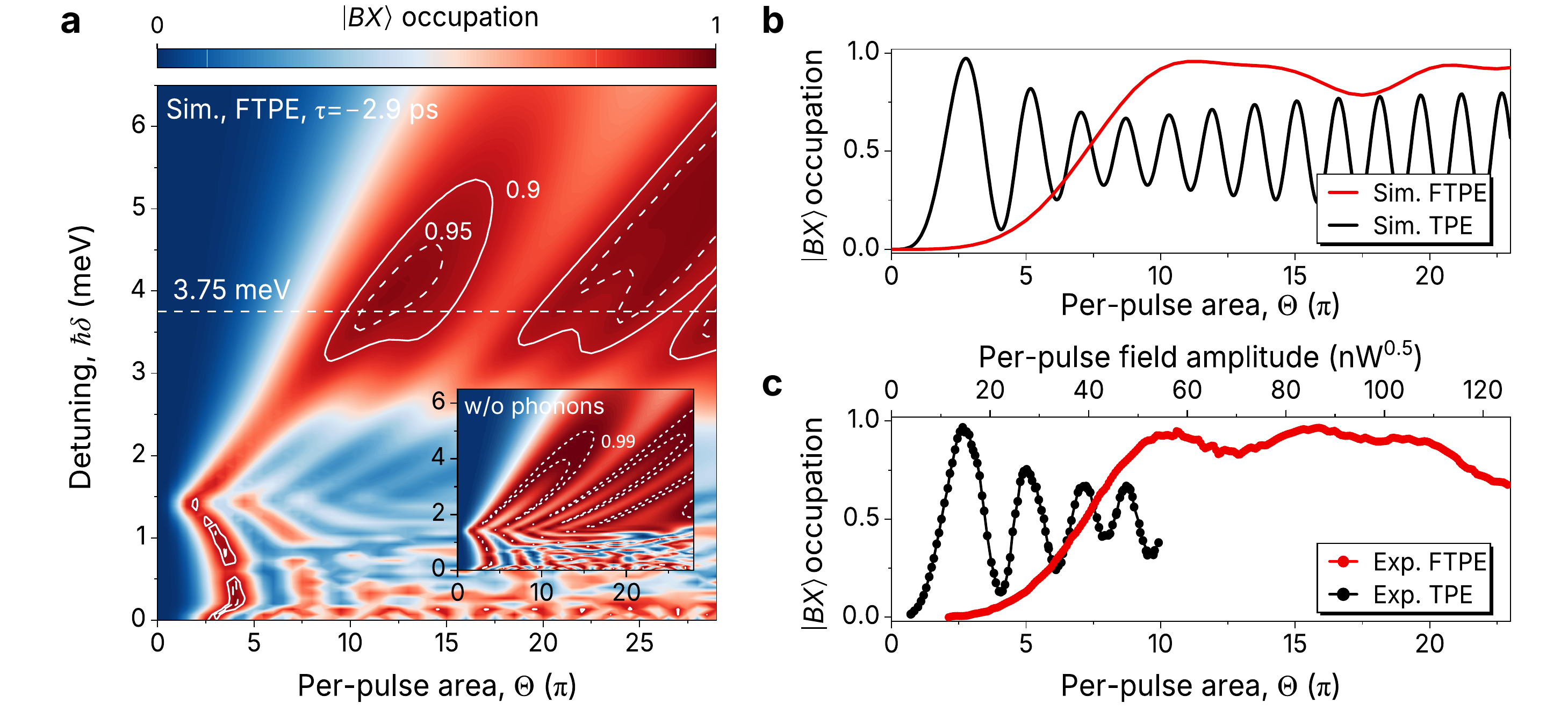}
 	\caption{\textbf{Deterministic and robust biexciton state preparation via FTPE.}  \textbf{a}, Calculated $BX$ occupation versus per-pulse area $\Theta$ and detuning $\delta$ with $\tau=-2.9$~ps. The horizontal line is where $\hbar \delta=3.75$~meV as used in \textbf{b} and \textbf{c}. Inset: calculated values without phonon interaction. \textbf{b}, Theoretically calculated $BX$ occupation as a function of the area of each pulse $\Theta$, where exciton-phonon interaction and radiative decay are taken into account. \textbf{c}, Measured intensity of $BX$ emission as a function of $\Theta$ under TPE and FTPE. The experimental $BX$ occupation and per-pulse area are normalized with two scaling factors by fitting the TPE curve to simulations (see details in Supplementary Fig.~\ref{Sfig:TPE rabi fit}).}
	\label{Rabi}
\end{figure*}

The working principle of our FTPE scheme can be understood by Floquet theory with respect to the period $T=2\pi/\delta$.
For detunings $\delta$ large enough so that $T\ll\tau_0$, fast oscillations can be eliminated and $H$ is replaced by a stroboscopic Hamiltonian $H^\textrm{eff.}$. %
If moreover $\delta$ is the dominant energy scale in the Hamiltonian, $\delta\gg E_b/\hbar, f$, the effective Hamiltonian $H^\textrm{eff.}$ can be obtained by systematic expansion in $1/\delta$~\cite{Bukov2015, Takegoshi2015}. This procedure yields a term that directly couples the ground to the biexciton state. The residual impact of the excitonic state is further eliminated by a Schrieffer-Wolff transformation, and we arrive at an effective Hamiltonian involving only ground and biexciton-like states with slowly varying driving envelopes. The details of this calculation can be found in the methods section. Formally, such a two-level system can be regarded as a spin precessing in a fictitious time-dependent effective magnetic field $B(t)$, which is described by the Hamiltonian
\begin{equation} \label{eq:stroboscopicHamiltonian_mainText}
    H^{\text{eff.}}(t) = \frac{\hbar}2 \big( B_x(t) \sigma_x + B_z(t) \sigma_z\big).
\end{equation}
Here \(\sigma_x\) and \(\sigma_z\) are Pauli matrices acting in the reduced two-level subspace spanned by the ground and biexciton states. The \(\sigma_x\) describes the effective coupling between these states, while the \(\sigma_z\) represents their effective detuning.

The effective fields are shaped by a combination of parameters of the exciting lasers, the biexciton binding energy and the detuning. 
In the special case of zero delay $\tau=0$ between the pulses [Fig.~\ref{Fig2}(a) and (c)], we find $B_z(t)=0$ and thus a Rabi-like dynamics with an effective pulse area $\int_{-\infty}^t B_x(t')dt'$ in the two-level system, which is a nonlinear function of the pulse area $\Theta$ of the original excitation pulses. Interestingly, to leading order, the effective pulse area $\Lambda$ scales as $\sim \Theta^2$ instead of the usual linear scaling (cf. Eq.~\eqref{eq:approxEffPulseArea} in the Methods Section). For finite delay $\tau$, $B_z(t)$ becomes nonzero [Fig.~\ref{Fig2}(b) and (d)], which facilitates the engineering of more complex dynamics. 

The time evolution of ground, exciton, and biexciton occupations is shown in Fig.~\ref{evolution}(e) and (f) for FTPE without and with time delay, respectively, at the optimal pulse area ($\Theta_{\mathrm{opt.}}$), which is defined as the pulse area corresponding to the first maximum of the final $BX$ occupation. The solid faded lines correspond to the solution of the full Hamiltonian Eq.~\eqref{eq:originalHamiltonian}, whereas the dashed lines describe the evolution according to the stroboscopic model of Eq.~\eqref{eq:stroboscopicHamiltonian_mainText}. 
Except for small-amplitude oscillations, the full numerical solution is well reproduced by the effective stroboscopic model. Thus, the behaviour of the physical system is indeed governed by the engineered effective magnetic field components $B_x(t)$ and $B_z(t)$. Finally, Fig.~\ref{evolution}(g) and (h) show the respective behaviour on the Bloch sphere, of which the south pole corresponds to the ground state $\left|0\right>$ and the north pole to the $BX$ state $\left|BX\right>$. 
To analyze the robustness of FTPE, we consider not only the optimal driving strengths $\Theta_{\text{opt.}}$ but also driving with increased pulse areas $1.25\Theta_{\text{opt.}}$ by scaling the envelopes $f(t)$ correspondingly.
Without delay, $\tau=0$, (left panel) the Bloch sphere trajectories run along a great circle (due to $B_z=0$). Thus, if the driving strength deviates from $\Theta_{\text{opt.}}$, the $BX$ population decreases maximally, as the great circle describes the straightest line on the Bloch sphere. 
With finite delay $\tau=-$2.9~ps, the component $B_z(t)$ is found to be antisymmetric in time. This component twists the trajectories out of the $yz$-plane. Nevertheless, the $BX$ state is reached for pulse area $\Theta_{\text{opt.}}$. Further increasing the driving strength not only increases the off-diagonal elements $B_x$ of the stroboscopic model but also the effective detuning $B_z$. The enhanced twisting leads to an elongation of the trajectory from the ground to the $BX$ state, such that overshooting, as found for $\tau=0$~ps, is compensated. This mechanism is responsible for the significantly enhanced robustness found for certain delays~$\tau$.

\addtocontents{toc}{\SkipTocEntry}
\subsection*{Experimental implementation}

To verify the experimental feasibility of the proposed FTPE scheme, we carry out the experiment with a single neutral droplet-etched GaAs/AlGaAs QD~\cite{Kersting2025}. The sample is placed in a closed-cycle cryostat and cooled to 3.6~K. Transform-limited Gaussian pulses are generated from a femtosecond laser and stretched into two tunable 12.49-ps pulses using 4-$f$ pulse shapers. The time delay $\tau$ between blue and red pulses is controlled by an optical delay line with a resolution of $\sim$15~fs. The exciton fine-structure splitting (FSS)~\cite{Bayer2002c} due to the QD's natural anisotropy is effectively eliminated by a continuous-wave (CW) laser via the ac Stark effect~\cite{Chen2024}. As considered in the theoretical model, the polarizations of lasers are all aligned with the QD's V dipole orientation, thereby minimizing coupling with the H-polarized exciton $\left|X_\text{H}\right>$. More experimental details are provided in the Supplementary Section~\ref{SI_Sec_setup}. 

\begin{figure*}[htbp]
    \centering
    \includegraphics[width=0.85\linewidth]{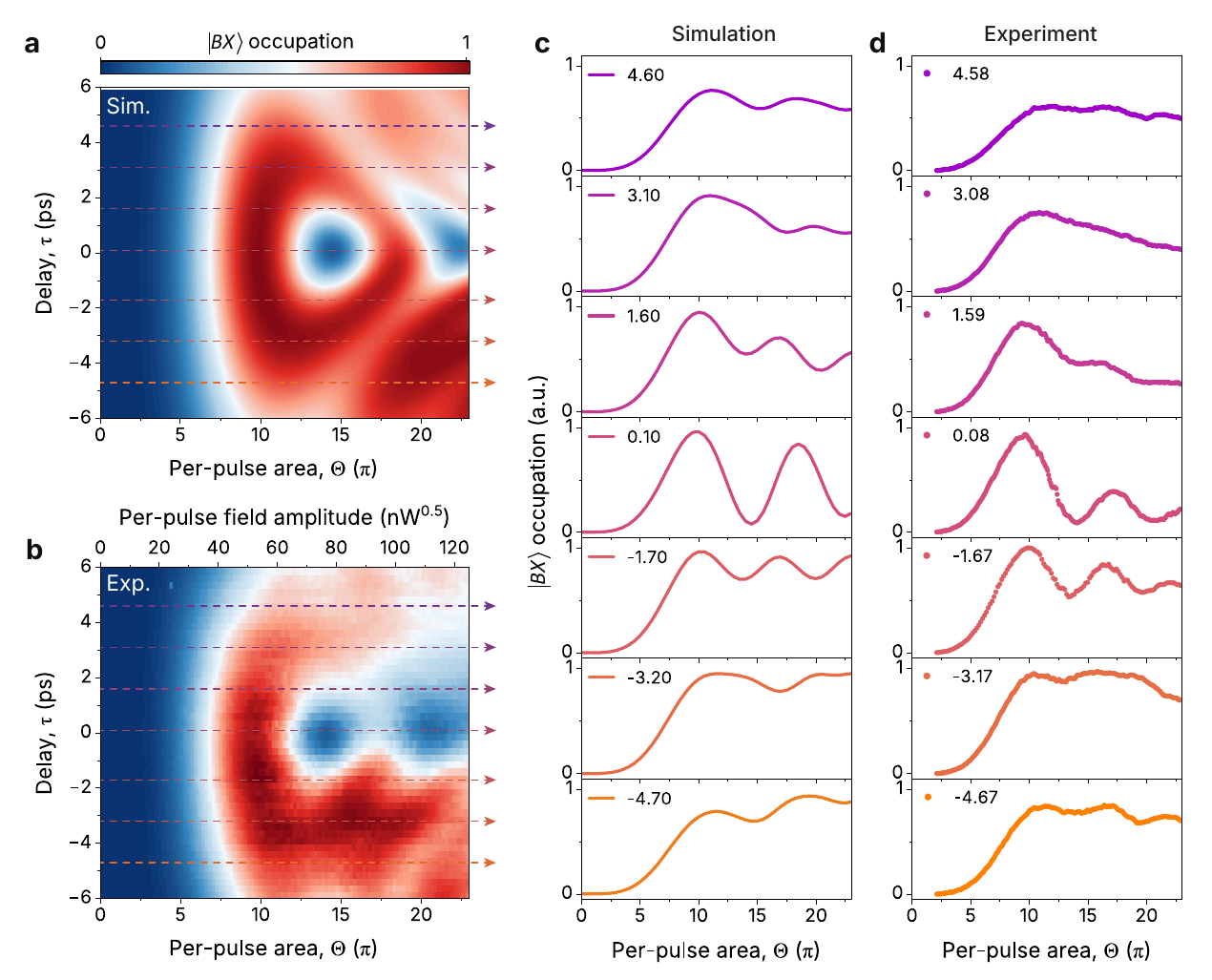}
    \caption{\textbf{Pulse delay dependence of FTPE.}
    \textbf{a} and \textbf{b}, Two-dimensional maps of the calculated (\textbf{a}) and measured (\textbf{b}) $BX$ occupation as functions of the per-pulse area $\Theta$ and pulse delay $\tau$, with $\hbar\delta=3.75~\mathrm{meV}$.
    \textbf{c}, Line cuts extracted from the calculated map in \textbf{a}, showing the calculated $BX$ occupation as a function of the per-pulse area for selected pulse delays ranging from $4.60$ to $-4.70$~ps.
    \textbf{d}, Line cuts extracted from the measured map in \textbf{b}, showing the measured $BX$ occupation as a function of the per-pulse area for selected pulse delays ranging from $4.58$ to $-4.67$~ps.}
    \label{Sfig:delay}
\end{figure*}

\addtocontents{toc}{\SkipTocEntry}
\subsection*{Robust biexciton preparation using FTPE}

We first perform a two-dimensional simulation by scanning over the laser pulse area $\Theta$ and detuning $\delta$ to map the $BX$ state preparation dynamics (Fig.~\ref{Rabi}a). This allows us to identify parameter regimes that enable efficient and robust $BX$ state preparation.
Qualitatively different parameter regimes can be identified: For TPE with small detunings, biexciton population is found at certain spots but the results are very sensitive to parameter fluctuations. At larger detunings, cusps at $\hbar\delta=E_b/2$ indicate the simultaneous resonance of the lasers with the $\left|0\right>$ to $\left|X_V\right>$ and $\left|X_V\right>$ to $\left|BX\right>$ transitions. Increasing the detuning further to about $\hbar\delta\approx E_b$, one reaches the FTPE regime. We note that low-order Magnus expansion captures only the FTPE regime well. The stroboscopic Floquet Hamiltonian describes the dynamics appropriately, down to small detunings where the pulse envelope no longer changes slowly with respect to $1/\delta$ (see Supplementary Section VII).
Notably, TPE theory~\cite{Stufler2006} involves an adiabatic approximation that breaks down for the large detunings required for FTPE (see Supplementary Section~\ref{sec:SM_tpe}).

The FTPE results show that near-unity preparation efficiency is achievable despite the presence of phonons, and high efficiency ($>90\%$) can be maintained across a wide range of excitation conditions, demonstrating the robustness of the scheme. 
FTPE can also be robust against spectral shifts, as we show in Supplementary Section~\ref{sec:SM_spectralShifts}. The theoretical model is described in the Methods Section, and it includes the full time-dependent Hamiltonian, radiative decay, as well as non-Markovian phonon effects, which are captured within the numerically exact process tensor framework ACE~\cite{Cygorek2024, CygorekPRX}. The inset of Fig.~\ref{Rabi}(a) shows calculated results without phonon effects, confirming that phonon assistance is not essential for $BX$ preparation. We attribute detrimental effects due to phonon coupling to phonon-assisted population of the exciton state, which are significant for detunings below about 3~meV, where the coupling to phonons is strongest. Choosing a larger detuning is therefore advisable to decouple from phonon effects. Without phonons, the reachable biexciton occupations are even higher, making the scheme even more attractive for emitters not affected by phonons, such as atomic systems.

Having identified the suitable laser pulse parameters, we next perform an experimental demonstration of FTPE. 
Fig.~\ref{Rabi}(c) shows experimentally measured $BX$ emission (red dots) obtained under FTPE with~$\hbar\delta=3.75$~meV. With increasing laser pulse area, the $BX$ emission exhibits a gradual increase followed by a slightly fluctuating plateau. As a comparison, the $BX$ emission (black dots) obtained under conventional TPE shows the well-known Rabi oscillation. The main damping source of the Rabi oscillations comes from the QD-phonon coupling~\cite{Ramsay2010d}. The experimentally achieved occupations and pulse area are calibrated by fitting the TPE Rabi oscillation data to numerical results with two scaling factors, as shown in Supplementary Fig.~\ref{Sfig:TPE rabi fit}. Both excitation schemes are well reproduced by our model [Fig.~\ref{Rabi}(b)]. As shown in Fig.~\ref{Rabi}(c), the first maximum of the $BX$ occupation under TPE (black dots, 96.8\%) occurs at a per-pulse area of 2.7$\pi$. The maximum $BX$ occupation achieved under FTPE is 96.6\% and appears at a per-pulse area of 15.9$\pi$. Importantly, the achieved final occupation shows significantly reduced sensitivity to variations in laser amplitude. This robustness to experimental fluctuations underscores the advantages of the FTPE scheme for practical quantum information applications. The small discrepancy between the theoretical and experimental peak occupations may arise from the uncertainty in the effective phonon-coupling coefficient extraction, as discussed in Supplementary Section~\ref{Ssec:Calibration}. The slight decrease of the experimental curve in the high-power regime may be related to additional excited-state channels, such as hole-state excitation (Supplementary Section~\ref{Ssec:ple}), that are not included in the present model.

The temporal separation between the two pulses plays a key role in achieving the robustness of FTPE. To comprehensively investigate the influence of this delay on FTPE performance, we simulate and measure the dependence of $BX$ occupation on the pulse area for various two-pulse delays. Figure~\ref{Sfig:delay}(a) presents a simulated two-dimensional colour map of $BX$ occupation as a function of pulse area and time delay. The detuning $\hbar\delta$ is fixed at 3.75 meV, consistent with the experimental condition. The asymmetry with respect to zero delay observed in the colour map arises from the LA-phonon-assisted excitation of the exciton~\cite{Quilter2015a}. For delays $\tau>0$, the blue pulse arrives first, populating the exciton state and thereby hindering the two-photon excitation process. A simulation excluding phonon effects is shown in Supplementary Fig.~\ref{Sfig:delay_dep}, where the delay dependence recovers a symmetric pattern. The corresponding experimentally measured two-dimensional map is presented in Fig.~\ref{Sfig:delay}(b), which reproduces the main features predicted by the simulation. Figures~\ref{Sfig:delay}(c) and \ref{Sfig:delay}(d) show line cuts taken from Figs.~\ref{Sfig:delay}(a) and \ref{Sfig:delay}(b), respectively, at selected delays. For \(\tau < 0\), the red-detuned pulse arrives first, such that competition from phonon-assisted excitation becomes negligible. In contrast to the pronounced Rabi-oscillation behavior observed near \(\tau = 0\) ps, the \(BX\) occupation becomes more robust against variations in pulse area over a finite range of negative delay. From the experimental results in Fig.~\ref{Sfig:delay}(d), we identify \(\tau=-2.5\) to \(-4.2\) ps as the experimentally optimal delay range.

\begin{figure}[tb]
\refstepcounter{fig}
	\includegraphics[width=1\linewidth]{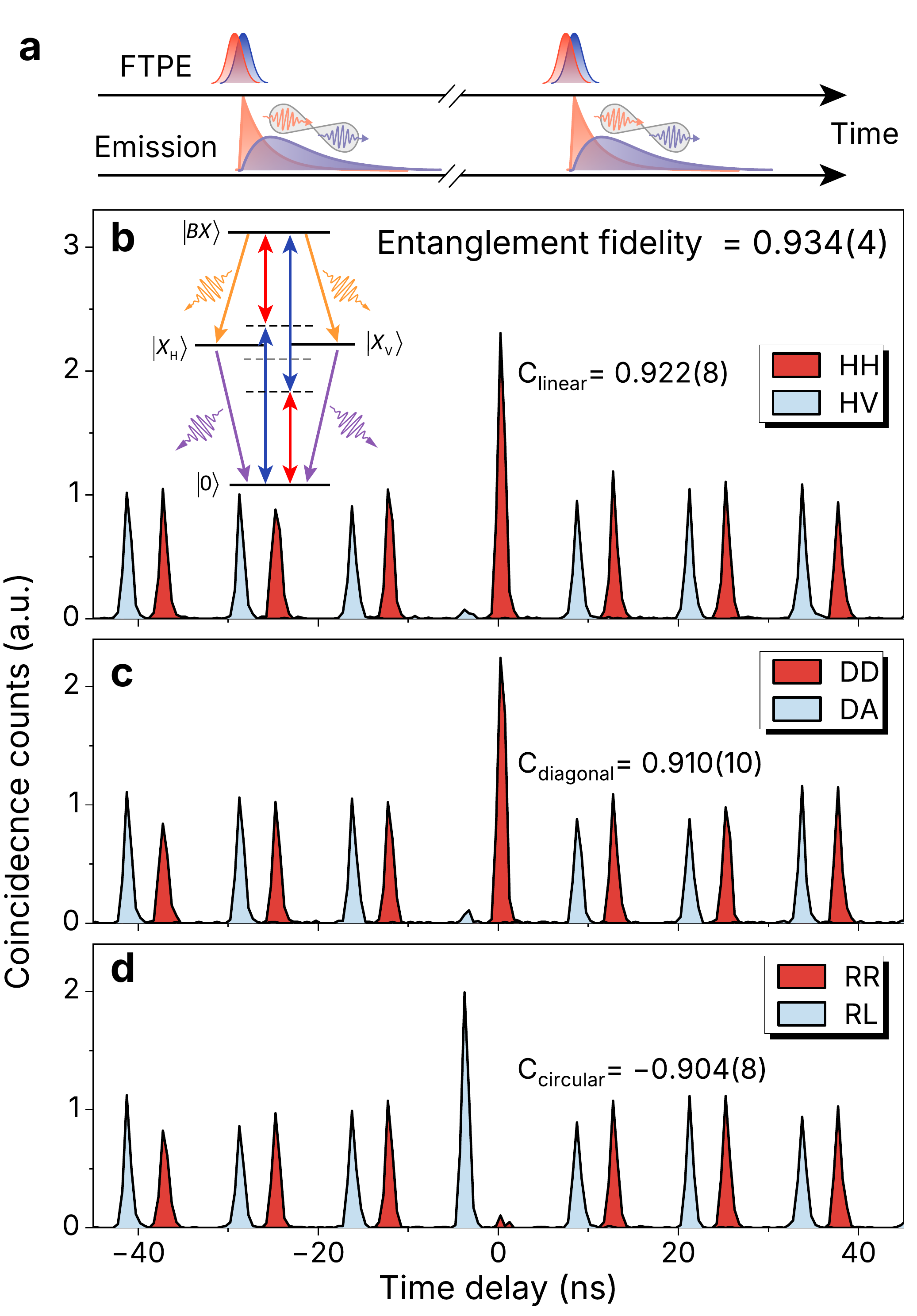}
    \caption{\textbf{Entanglement characterization under FTPE.} \textbf{a}, Schematic of polarization entangled photon pairs generated by FTPE. \textbf{b}, Detected $BX$-$X$ photon cross-correlation coincidence counts under FTPE in linear basis (H, horizontal; V, vertical). Inset: The cascade radiative recombination of $BX$ state provides two polarization-entangled photons. \textbf{c}, In diagonal basis (D, diagonal; A, anti-diagonal). \textbf{d}, In circular basis (R, right circular; L, left circular).}
	\label{entangle}
\end{figure}
\addtocontents{toc}{\SkipTocEntry}
\subsection*{On-demand generation of entangled photons}

One promising application of our FTPE scheme is to deterministically generate entangled photon pairs. To assess the practicality, it is crucial to examine both the entanglement fidelity of the FTPE-generated photon pairs and the compatibility of the FTPE scheme with experimental techniques required for the collection and verification of entangled photons.
To this end, we investigate the cascade emission from the $BX$ state prepared by FTPE, which is operated as a source of polarization-entangled photon pairs [Fig.~\ref{entangle}(a) and (b) inset]. Since the FTPE pulses are far detuned from $BX$ and $X$ photon emission peaks, they can be removed simply by spectral filters without affecting the polarization state of entangled photons. The entanglement fidelity is then characterized by separating $BX$ and $X$ photons using a polarization-insensitive grating, followed by the measurement of the polarization-dependent second-order correlation functions. The intrinsic FSS of \(3.4(2)~\upmu\mathrm{eV}\) is eliminated by the ac Stark effect induced by the 13.5~\(\upmu\mathrm{W}\) CW laser with a red detuning of 300~\(\upmu\mathrm{eV}\)~\cite{Chen2024}. Figure~\ref{entangle} shows six cross-correlation measurements in the rectilinear [HH and HV, Fig.~\ref{entangle}(b)], diagonal [DD and DA, Fig.~\ref{entangle}(c)], and circular [RR and RL, Fig.~\ref{entangle}(d)] polarization bases under FTPE, with each histogram composed of 500-ps time bins. By analyzing the degrees of polarization correlation ($C_\mu$, where $\mu$ is the polarization basis), the fidelity to the Bell state $\left|\psi^{+}\right\rangle=\frac{1}{\sqrt{2}}\left(\left|H_{B X}\right\rangle\left|H_X\right\rangle+\left|V_{B X}\right\rangle\left|V_X\right\rangle\right)$ is obtained as: 
\begin{equation}
f=\frac{1+C_{\text {linear }}+C_{\text {diagonal }}-C_{\text {circular }}}{4}=0.934(4).
\end{equation}
The uncertainties are calculated using error propagation, assuming Poisson statistics for the coincidence counts. In the idealized calculation without FSS and the CW laser, the fidelity reaches 97.7\% (Supplementary Section~\ref{Ssec:simulation fidelity}), with the remaining 2.3 percentage-point reduction from unity caused by the pulse-induced transient spectral shift. Including the CW laser reduces the calculated fidelity further to 95.3\%, close to the experimentally measured value of 93.4\%. This additional reduction is associated with CW-induced state dressing, which is also supported by the power dependence of the autocorrelation measurements (see Supplementary Section~\ref{Ssec:cw fss}). Importantly, the FTPE protocol is compatible with other established FSS-erasure techniques, such as strain~\cite{Huber2018b}, magnetic-field~\cite{Stevenson2006a}, and static electric-field~\cite{Bennett2010} tuning, which could avoid the associated state dressing and thereby enable higher entanglement fidelity. These results demonstrate that the FTPE protocol is a reliable method for generating highly entangled photon pairs.

\addtocontents{toc}{\SkipTocEntry}
\section*{Discussion and conclusion}

We have demonstrated the design and implementation of a robust biexciton state preparation scheme for a semiconductor quantum dot based on our concept of ultrafast all-optical Floquet engineering. Appropriately choosing the parameters of two differently coloured laser pulses lets us arrange a situation where the quantum dot experiences periodic driving within the laser duration. This allows us to engineer a stroboscopic Hamiltonian connecting the initial state with the target state of the protocol and to control the effective time-dependent driving between them.
Thereby, we can identify a suitable parameter regime where FTPE is robust against fluctuations in the laser power and where detrimental phonon effects are also of minor importance. In particular, we have identified the delay between the two pulses to be a pivotal tuning knob for reaching the robust regime.
The theoretical predictions are well reproduced in our experiments. 

The selectable laser detuning of FTPE can provide greater experimental flexibility and relax the demand for laser filtering. In the present weakly confined GaAs/AlGaAs QD, the practical detuning window is finite but remains sizable (\(\gtrsim 1~\mathrm{meV}\), Supplementary Section~\ref{Ssec:ple}). This restriction is specific to the present QD system rather than an intrinsic limitation of FTPE.
Moreover, the high entanglement fidelity achieved suggests that the FTPE scheme does not induce additional dephasing channels, further validating its practicality for quantum information applications. Notably, the laser pulse parameters used in this work, such as duration, waveform, and chirp, can be further optimized. Future improvements could be realized through quantum optimal control~\cite{Werschnik2007} by systematically tailoring these parameters to enhance the efficiency and robustness of state preparation.

In Table~\ref{tab:table1}, we provide a comparison of FTPE with representative excitation schemes used for entangled-photon generation in QDs. FTPE offers a highly robust and filtering-friendly route for generating entangled photon pairs, while maintaining high fidelity and preparation efficiency comparable to those of conventional TPE. As also detailed in the Supplementary Section~\ref{diff_to_others}, the proposed FTPE scheme is distinct from other excitation schemes: the parameter regime for FTPE with detuning $\delta > E_b/(2\hbar)$ is opposite to the limit of resonant TPE, which can be generalized to small but nonzero detuning $\delta \ll E_b/(2\hbar)$. FTPE works almost the same when the binding energy is negative. As binding energy decreases, more laser power is needed and for zero binding energy no robust excitation can be obtained. Moreover, although robustness is also a feature found in adiabatic protocols, especially in the most robust region of finite pulse delays, FTPE is not adiabatic on an instantaneous eigenstate basis. The optical selection rules differ from stimulated Raman adiabatic passage (STIRAP), and the resonance conditions of FTPE differ from those in other two-colour excitation schemes such as SUPER. These unique characteristics collectively highlight the distinctions of FTPE. Hence, the concept of all-optical single-shot Floquet driving provides a basis for developing platform-tailored excitation protocols for quantum emitters with suitable ladder-type structure, whenever one wants to drive transitions between states that are only indirectly coupled via off-resonant intermediate states. The essential requirements are compatible optical selection rules, sufficiently large but experimentally accessible detunings to suppress real occupation of the intermediate state while retaining effective coupling, and coherence times long compared with the single-shot driving window. Beyond the semiconductor QD system demonstrated here, this concept may also be applicable to selected molecular~\cite{Ahmed2006}, atomic~\cite{Srivathsan2013}, colour-center~\cite{Golter2014a}, and superconducting artificial-atom systems~\cite{Liu2015}.

\begin{table*}
\caption{Comparison of existing \textit{BX} preparation schemes in semiconductor QDs}
\centering
\footnotesize
\renewcommand{\arraystretch}{1.2}
\setlength{\tabcolsep}{7pt}
\begin{tabular}{ccccccc}
\hline\hline
\multirow{2}{*}{Scheme}
& \multicolumn{2}{c}{Preparation efficiency}
& \multicolumn{1}{c}{Entanglement fidelity}
& \multirow{2}{*}{Ref.}
& \multirow{2}{*}{\makecell{Pulse-area\\robustness}}
& \multirow{2}{*}{\makecell{Filtering\\difficulty}} \\
\cline{2-4}
& Theory & Experiment & Experiment & & & \\
\hline

\multirow{3}{*}{TPE}
& $>99\%$ without phonons & $\sim82\%$ & -- & \cite{Stufler2006} & \multirow{3}{*}{Weak} & \multirow{3}{*}{Moderate} \\
\cline{2-5}
& -- & $\sim90\%$ & $88\%$ & \cite{Liu2019e} & & \\
\cline{2-5}
& -- & n.q. & $97.8\%$ & \cite{Huber2018b} & & \\
\hline

\multirow{4}{*}{\makecell{Phonon-assisted\\TPE}}
& $>99\%$ & -- & -- & \cite{Glassl2013c} & \multirow{4}{*}{Strong} & \multirow{4}{*}{Difficult} \\
\cline{2-5}
& -- & $80\%$ & $90\%$ & \cite{Reindl2017a} & & \\
\cline{2-5}
& -- & $95\%$ & -- & \cite{Bounouar2015b} & & \\
\cline{2-5}
& -- & $>80\%$ & -- & \cite{Ardelt2014a} & & \\
\hline

\multirow{2}{*}{ARP}
& \begin{tabular}{c} $>90\%$ at 4 K \\ $>98\%$ at 1 K \end{tabular} & -- & -- & \cite{Glassl2013b} & \multirow{2}{*}{Strong} & \multirow{2}{*}{Difficult} \\
\cline{2-5}
& -- & $>95\%$ & -- & \cite{Kappe2023} & & \\
\hline

\multirow{3}{*}{SUPER}
& $>99\%$ without phonons & -- & -- & \cite{Bracht2023a} & \multirow{3}{*}{Moderate} & \multirow{3}{*}{Easy} \\
\cline{2-5}
& $>95\%$ at 4 K & -- & -- & \cite{Bracht2023} & & \\
\cline{2-5}
& -- & n.q. & -- & \cite{Piccinini2025} & & \\
\hline

FTPE
& \makecell{$>99\%$ without phonons \\ $96.7\%$ at 3.6 K} & $96.6\%$ & $93.4\%$ $(97.7\%)^{a}$ & This work & Strong & Easy \\
\hline\hline

\multicolumn{7}{l}{$^{a}$ Ideal fidelity without FSS and CW laser. n.q., not quantified. --, not reported.}
\end{tabular}
\label{tab:table1}
\end{table*}

\newpage
\addtocontents{toc}{\SkipTocEntry}
\section*{Methods}

\addtocontents{toc}{\SkipTocEntry}
\subsection*{Numerical methods}
In our simulations, we include longitudinal acoustic (LA) phonons. Their coupling to the QD is included using the Hamiltonian
\begin{equation}
H_{\mathbf{ph}}/\hbar=\sum_{\mathbf{q}} \omega_{\mathbf{q}} b_{\mathbf{q}}^{\dagger} b_{\mathbf{q}}+\sum_{\mathbf{q},\nu}n_\nu\left(g_{\mathbf{q}} b_{\mathbf{q}}+g_{\mathbf{q}}^* b_{\mathbf{q}}^{\dagger}\right)|\nu\rangle\langle\nu|,
\end{equation}
where $b_{\mathbf{q}}^{\dagger}$ $(b_{\mathbf{q}})$ is the creation (annihilation) operator for a LA phonon with wave vector $\mathbf{q}$ and frequency $\omega_{\mathbf{q}}$, $|\nu\rangle \in\left\{\left|X_V\right\rangle,\left|BX\right\rangle\right\}$ is the excitonic state and $n_\nu$ is the number of excitons present in the state $|\nu\rangle$. $g_{\mathbf{q}}$ denotes the coupling constant, which depends on crystal mass density, mode volume, material deformation potential and so on. We use these material parameters taken from the literature~\cite{Krummheuer2005} and leave the electron (hole) wavefunction radius as the only free parameter, where the wavefunctions correspond to a harmonic confinement potential.

The radius of the electronic wavefunction is determined by fitting the calculated pulse-area dependence of the $|BX\rangle$-occupation for TPE to the experimental TPE results, which are displayed in Fig.~\ref{Rabi}~b) and c). We obtain the radius $a_e = 5.5$~nm. For the corresponding hole radius, we use $a_h = a_e/1.15$. All subsequent FTPE calculations were done using these values.

We solve the combined QD-phonon Hamiltonian using a state-of-the-art process tensor method \cite{JoergensenPollock2019, Cygorek2022, Cygorek2024}. This is necessary to accurately include phonons because it allows for short time-step lengths, which is necessary in this case due to the potentially large detunings $\delta$. Other numerically complete methods, in particular, path-integral methods~\cite{Vagov2011} %,Barth2016,Cygorek2017, 
would demand unfeasibly large amounts of memory.
\addtocontents{toc}{\SkipTocEntry}
\subsection*{Derivation of the effective Hamiltonian} \label{subsec:DerivEffHamilton}
Here, we summarize the mapping from the two-colour two-photon excitation of the three-level system composed of ground, exciton and biexciton states to an effective two-level system consisting predominantly of ground and biexciton states.
We focus on the regime where the detuning $\delta$ is the dominant frequency scale. Then, the envelopes $f(t\pm\tau/2)$ in the driving term
$\Omega(t)=f(t-\tau/2)e^{-i\delta (t-\tau/2)}+f(t+\tau/2)e^{i\delta (t+\tau/2)}$ vary slowly during a single oscillation period $T=2\pi/\delta$.

To separate the time scales, we first assume the envelopes $f(t)\approx f$ to be constant 
over time $T$.  
We seek a stroboscopic Hamiltonian $\bar{H}$, which is similarly constant
over short times $t\lesssim T$, but reproduces the time evolution over a single oscillation period $T$. It is defined by the condition
\begin{align}
U(T)=e^{-\frac{i}{\hbar} \bar{H} T} 
\stackrel{!}{=} \mathcal{T}e^{-\frac{i}{\hbar} \int\limits_0^{T} H(t) dt},
\label{eq:Uequals}
\end{align}
where $\mathcal{T}$ denotes the time-ordering operator. The stroboscopic Hamiltonian $\bar{H}=\bar{H}(f)$ parametrically depends on the pulse envelope $f$.

Thus, the dependence of $\bar{H}$ on a coarse-grained time $t$ can be reintroduced via the parametric dependence  $\bar{H}(t)=\bar{H}(f(t))$. 

While it is straightforward to obtain $\bar{H}$ for a fixed value of $f$ numerically, 
analytic insights can be gained by considering the
Magnus expansion $\bar{H}=\sum_{n=0}^\infty \bar{H}^{(n)}$. The leading-order terms are given by~\cite{Takegoshi2015}
\begin{subequations}
\begin{align}
\bar{H}^{(0)}=&\frac{1}{T} \int\limits_0^T dt_1\,H(t_1) \\
\bar{H}^{(1)}=&\frac{-i}{2 \hbar T}
\int\limits_0^T dt_1\int\limits_0^{t_1} dt_2 \,[H(t_1),H(t_2)],
\end{align}
\begin{align}
\bar{H}^{(2)}=&\frac{-1}{6\hbar^2 T} \int\limits_0^T dt_1
\int\limits_0^{t_1} dt_2 \int\limits_0^{t_2} dt_3
\,\Big(\big[H(t_1),[H(t_2),H(t_3)]\big]
\nonumber\\&+\big[H(t_3),[H(t_2),H(t_1)]\big]\Big),\\
\bar{H}^{(3)}=&\frac{i}{12\hbar^3 T} \int\limits_0^T dt_1
\int\limits_0^{t_1} dt_2 \int\limits_0^{t_2} dt_3\int\limits_0^{t_3} dt_4
\nonumber\\&
\bigg(\Big[\big[[H(t_1),H(t_2)],H(t_3)\big],H(t_4)\Big]
\nonumber\\&+\Big[H(t_1),\big[[H(t_2),H(t_3)],H(t_4)\big]\Big]
\nonumber\\&+\Big[H(t_1),\big[H(t_2),[H(t_3),H(t_4)]\big]\Big]
\nonumber\\&+\Big[H(t_2),\big[H(t_3),[H(t_4),H(t_1)]\big]\Big]
\bigg).
\end{align}
\end{subequations}
One finds that, to zeroth order, the dynamics on a coarse-grained timescale is governed
by the time-averaged Hamiltonian $\bar{H}^{(0)}$. Yet the commutators in the 
higher-order terms can lead to qualitatively different transitions that are not directly
visible from the original Hamiltonian $H(t)$. 

For example, if there is no delay between the two pulses, $\Omega(t)=2f(t)\cos(\delta t)$ and 
\begin{subequations} \label{eq:MagnusFloquetHamiltonian_Appendix}
\begin{align}
\bar{H}^{(0)}=&\left(\begin{array}{ccc}
0&0&0\\0& E_b/2&0 \\0&0&0
\end{array}\right),
\\
\bar{H}^{(1)}=&\,0,
\\
\bar{H}^{(2)}=&\,\frac{E_b f}{4\hbar\delta^2} 
\left[\hbar f\left(\begin{array}{ccc} 1&0&1\\0& -2&0 \\1&0&1 \end{array}\right)
-\frac{E_b}{2}\left(\begin{array}{ccc} 0&1&0\\1& 0&1 \\0&1&0 \end{array}\right)
\right].
\end{align}
\end{subequations}
The zero-order term ensures that the exciton state remains off-resonant by
energy $E_b/2$. Because $n$-th order terms scale as $1/\delta^n$ with the detuning
$\delta$, the zero-order term ensures that the excitonic state only becomes weakly
occupied during the dynamics in the regime of interest where $\delta$ dominates. 
The first-order term vanishes when there is no delay between the pulses.
Thus, the dynamics is mainly determined by the second-order term $\bar{H}^{(2)}$,
which itself consists of two contributions. The first provides corrections to
the diagonal energies but more importantly drives direct transitions between
the ground and the biexciton state with an effective Rabi frequency
$E_b f^2(t)/(2\hbar\delta^2)$. The second contribution
describes conventional transitions between ground and exciton and between exciton
and biexciton states. However, as the exciton state is strongly off-resonant, the latter 
processes remain largely virtual and only lead to small quantitative corrections
to the dynamics.

In fact, we find that the exciton state remains very weakly occupied even in the more complex case of finite delay between the pulses. This makes it possible to eliminate the exciton state altogether by a Schrieffer-Wolff transformation~\cite{SchriefferWolff,WinklerBook}. There, the basis is slightly rotated by a unitary transformation such that coupling terms between the third, predominantly excitonic state and the remaining two states, which are predominantly composed of the ground and biexciton state, vanish to leading order. To the second order, the effective two-level Hamiltonian is constructed from the stroboscopic Hamiltonian $\bar{H}$ by
\begin{align}
\label{eq:HSW}
H^\textrm{eff.}_{m,n}=& \bar{H}_{m,n} 
+\frac 12
\bigg(\frac{\bar{H}_{m,X_V}\bar{H}_{X_V,n}}{\bar{H}^{(0)}_{m,m}-\bar{H}^{(0)}_{X_V,X_V}}
+\frac{\bar{H}_{m,X_V}\bar{H}_{X_V,n}}{\bar{H}^{(0)}_{n,n}-\bar{H}^{(0)}_{X_V,X_V}}\bigg),
\end{align}
where $n,m\in\{BX,0\}$. Higher-order Schrieffer-Wolff contributions are doubly suppressed, on the one hand by factors $1/\delta^n$ in the numerator, which stems from higher-order Magnus expansion terms (valid for $\delta\gg E_b/(2\hbar))$, and on the other hand by factors $E_b^n$ in the Schrieffer-Wolff denominators (valid when $E_b$ is much larger than the prefactors in 
$\bar{H}^{(m)}$ for $m>0$). 
Doing this and introducing the time dependence of the laser envelopes, $f\to f(t)$, we obtain the effective Hamiltonian
$H^{\text{eff.}} = \frac{\hbar}2(B_x(t)\sigma_x +B_z(t)\sigma_z)$ with components $B_x(t)$ and $B_z(t)$ obtained by calculating the inner product
\begin{equation}
    B_{x/z}(t)=\frac{1}{\hbar}\text{Tr}[\sigma_{x/z} H^\text{eff.}].
\end{equation}

For the case without delay, we can now also derive analytical formulas for the approximate biexciton occupation for different pulse amplitudes, using the well-known expressions from resonant Rabi systems. As was seen in Fig.~\ref{evolution} of the main text, the diagonal elements of \eqref{eq:HSW} are zero for $\tau=0$~ps, which gives an effective pulse area of
\begin{equation}
    \label{eq:approxEffPulseArea}
    \Lambda \coloneqq \frac{2}{\hbar}\int_{-\infty}^\infty H^{\text{eff}}_{BX,0}(t)\,\text{d}t = \frac{E_b}{4\sqrt{\pi}\hbar\delta^2s}\left(1-\frac{E_b^2}{8\hbar^2\delta^2}\right)\Theta^2.
\end{equation}
Using this, it is possible to calculate the final biexciton occupation using
\begin{equation}
    |\langle BX|\psi(t=\infty)\rangle|^2 = \text{sin}^2\left(\frac{\Lambda}{2}\right).
\end{equation}
Importantly, the effective pulse area $\Lambda$ scales with the square of the original pulse area $\Theta$, which was used for the two pulses.
\\

\addtocontents{toc}{\SkipTocEntry}
\section*{Data availability}
The raw data that support the findings of this study are available on Zenodo at https://doi.org/10.5281/zenodo.21787444~\cite{Yan2026Zenodo}.

\addtocontents{toc}{\SkipTocEntry}
\section*{Code availability}
The code that has been used for this work is available from the corresponding authors upon request.

\addtocontents{toc}{\SkipTocEntry}

\end{bibunit}
\begin{bibunit}

\addtocontents{toc}{\SkipTocEntry}
\section*{Acknowledgements}
We thank Dr.~Doris~Reiter for fruitful discussions. We acknowledge the Micro/Nano Fabrication Center at Zhejiang University for its facility support and technical assistance. We thank the National Supercomputer Center in Guangzhou.

\addtocontents{toc}{\SkipTocEntry}
\section*{Author contributions}
F.L. and M.C. conceived the project. H.-G.B., A.D.W. and A.L. grew the wafer and fabricated the sample. J.-Y.Y. built the setup. J.-Y.Y., H.-Y.G. and F.-Y.L. carried out the experiments. J.-Y.Y., P.C.A.H., M.C. and F.L. analysed the data. P.C.A.H. and M.C. performed the theoretical modeling. W.E.I.S., B.W., Z.G., C.-Y.J., V.M.A., D.-W.W., M.C., and F.L. provided supervision and expertise. J.-Y.Y., P.C.A.H., M.C. and F.L. wrote the manuscript with comments and input from all the authors.

\addtocontents{toc}{\SkipTocEntry}
\section*{Funding}
F.L. acknowledges support from the National Key Research and Development Program of China (2023YFB2806000), Zhejiang Provincial Natural Science Foundation of China (No. LR25F040001), and Self-initiated Research Projects funded by the State Key Laboratory of Extreme Photonics and Instrumentation. H.-G.B., A.D.W., and A.L. acknowledge support by the BMBF Project QR.N 16KIS2200, QuantERA BMBF EQSOTIC 16KIS2061, DFG ML4Q EXC 2004/1, and the DFH/UFA, Project CDFA-05-06. M.C. is supported by the Return Program of the State of North Rhine-Westphalia. Open Access funding enabled and organized by Projekt DEAL.

\addtocontents{toc}{\SkipTocEntry}
\section*{Competing interests}
The authors declare no competing interests.

\addtocontents{toc}{\SkipTocEntry}
\section*{Additional information}
The online version contains supplementary material available at https://doi.org/10.1038/s41467-026-77221-9. Correspondence and requests for materials should be addressed to Moritz~Cygorek or Feng~Liu.

\addtocontents{toc}{\SkipTocEntry}

% No bibliography in this bibunit.

\end{bibunit}
\begin{bibunit}

\setcounter{section}{0}
\setcounter{equation}{0}
\setcounter{figure}{0}
\setcounter{table}{0}

\renewcommand\thesection{\Roman{section}}
\renewcommand{\figurename}{Supplementary Figure}
\renewcommand{\tablename}{Supplementary Table}
\renewcommand{\thetable}{\arabic{table}}

\renewcommand{\theequation}{S\arabic{equation}}
\renewcommand{\bibnumfmt}[1]{[S#1]} % This line requires natbib
\renewcommand{\citenumfont}[1]{S#1} % This line requires natbib

\onecolumngrid
\clearpage
\newpage

\begin{Large}
\begin{center}

\textbf{Supplementary information: Robust biexciton preparation for entangled photon-pair generation by single-shot Floquet driving}
\end{center}
\end{Large}

\setcounter{page}{1}
\begin{center}
Jun-Yong Yan$^*$, Paul C. A. Hagen$^*$, Hang-Yu Ge, Fang-Yuan Li, Hans-Georg Babin, Wei E. I. Sha, Bo Wang, Zhi-hua Gan, Andreas D. Wieck, Arne Ludwig, Chao-Yuan Jin, Vollrath Martin Axt, Da-Wei Wang, \\Moritz Cygorek$^\dagger$, and Feng Liu$^\ddag$
\end{center}

\begin{center}
\textbf{CONTENTS}
\end{center}

\noindent I. Experimental setup \dotfill 2\\
II. Biexciton preparation efficiency at different pulse delays without phonon \dotfill 3\\
III. Tuning QD FSS via ac Stark effect \dotfill 3\\
IV. Calibration of pulse area, occupation, and electronic wavefunction radius \dotfill 4\\
V. Parameters used in the calculations \dotfill 4\\
VI. Determination of suitable detuning for FTPE in GaAs QD \dotfill 5\\
VII. Validity of the Floquet model \dotfill 6\\
VIII. Theoretical entanglement fidelity \dotfill 7\\
IX. Robustness of FTPE against parameter variations \dotfill 8\\
\hspace*{1.5em}A. Influence of the pulse delay \dotfill 8\\
\hspace*{1.5em}B. Influence of spectral shifts \dotfill 9\\
\hspace*{1.5em}C. Influence of pulse-area imbalance \dotfill 9\\
X. Comparison with other excitation mechanisms \dotfill 11\\
\hspace*{1.5em}A. Nearly monochromatic two-photon resonant excitation \dotfill 11\\
\hspace*{1.5em}B. Adiabatic protocols \dotfill 11\\
\hspace*{1.5em}C. STIRAP \dotfill 12\\
\hspace*{1.5em}D. SUPER and dichromatic excitation \dotfill 13\\
XI. Relative phase difference between the lasers \dotfill 14\\
References \dotfill 15

\newpage
\section{Experimental setup} \label{SI_Sec_setup}
\label{sec:setup}

A schematic of the optical characterization setup is illustrated in Supplementary Fig.~\ref{setup}. The QD sample is placed in the 3.6-K cryostat (attocube). Transform-limited femtosecond pulses are generated by a Ti:sapphire oscillator (Coherent) and subsequently split and shaped into a pair of pulses with different colours (denoted blue and red) by two pulse shapers. The temporal delay between the blue and red pulses is controlled by an optical delay line (Newport). A tunable CW laser (M squared) is set to be slightly red-detuned with respect to the $\left|BX\right>\rightarrow\left|X_V\right>$ transition and used to tune the QD FSS. The polarization of both the pulsed and CW lasers is aligned with the V dipole orientation of the QD, ensuring minimal coupling to the H dipole. 

The emitted photons are collected by an aspheric objective lens (Thorlabs) and directed to a custom-built entanglement measurement setup. A filtering system, comprising a volume phase holographic transmission grating (Coherent) and notch filters (Optigrate), is employed to spectrally eliminate laser scattering and separate the BX and X photons. A set of wave plates (Thorlabs), including two QWPs and one HWP, is used to compensate for polarization misalignment between the QD emission and the tomography basis~\cite{Zhou2019,Wang2007}. Finally, the entangled photons are projected onto different polarization bases and then detected by two single-photon avalanche detectors (Excelitas) for cross-correlation measurements.

\begin{figure*}[htbp]
\refstepcounter{fig}
	\includegraphics[width=1\linewidth]{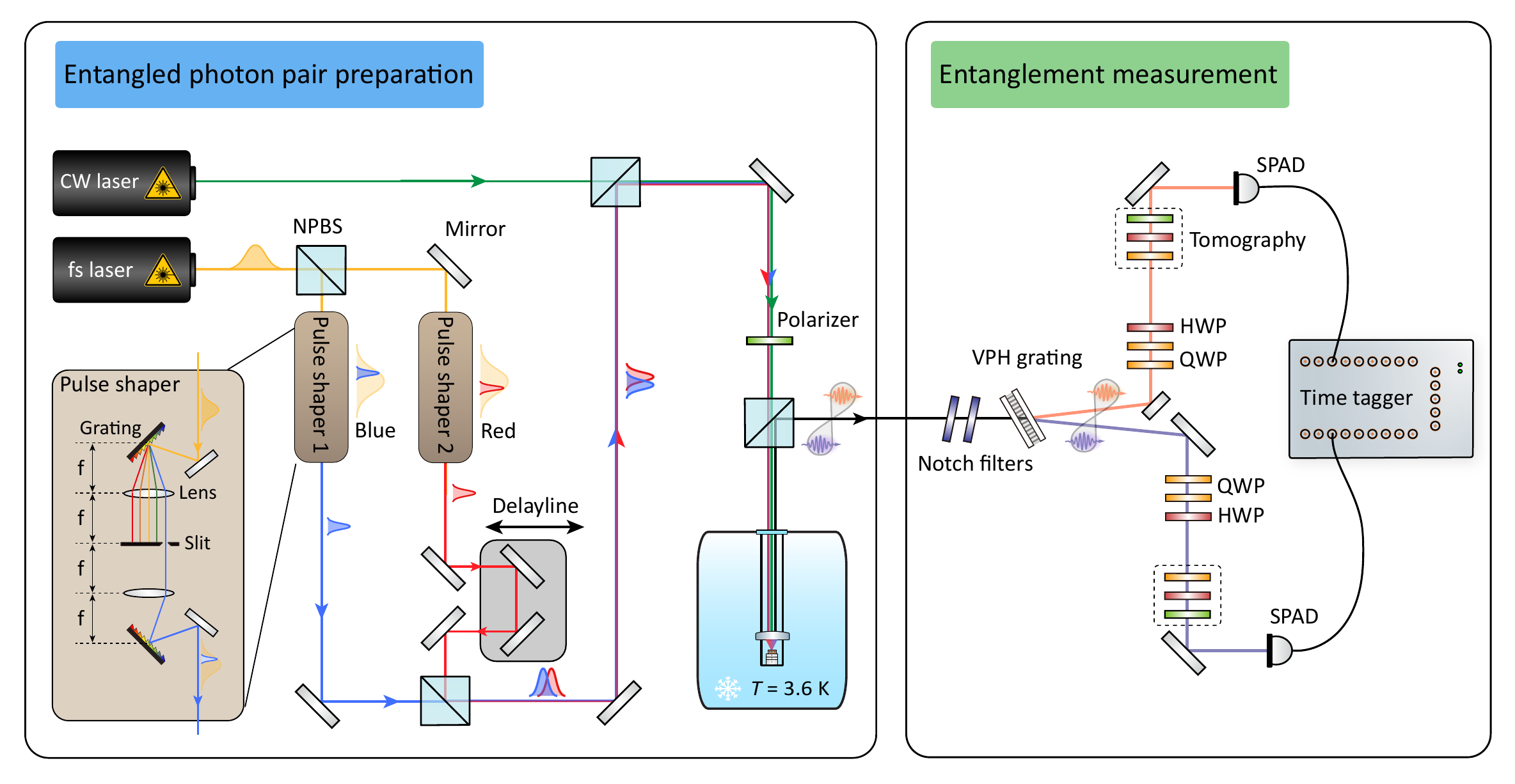}
        \caption{\textbf{Sketch of the experimental setup.} The setup consists of two main parts: entangled photon pair preparation and entanglement measurements. NPBS: non-polarizing beam splitter; VPH grating: volume phase holographic grating; HWP: half-wave plate; QWP: quarter-wave plate; SPAD: single-photon avalanche diode.}
	\label{setup}
\end{figure*}

\newpage
\section{Biexciton preparation efficiency at different pulse delays without phonon}
We also simulated the $BX$ occupation's dependence on pulse area at different pulse delays, without considering QD-phonon coupling. As shown in Supplementary Fig.~\ref{Sfig:delay_dep}, a symmetric two-dimensional pattern about delay $\tau=0$~ps is recovered.

\begin{figure}[htbp]
    \centering
    \includegraphics[width=0.4\linewidth]{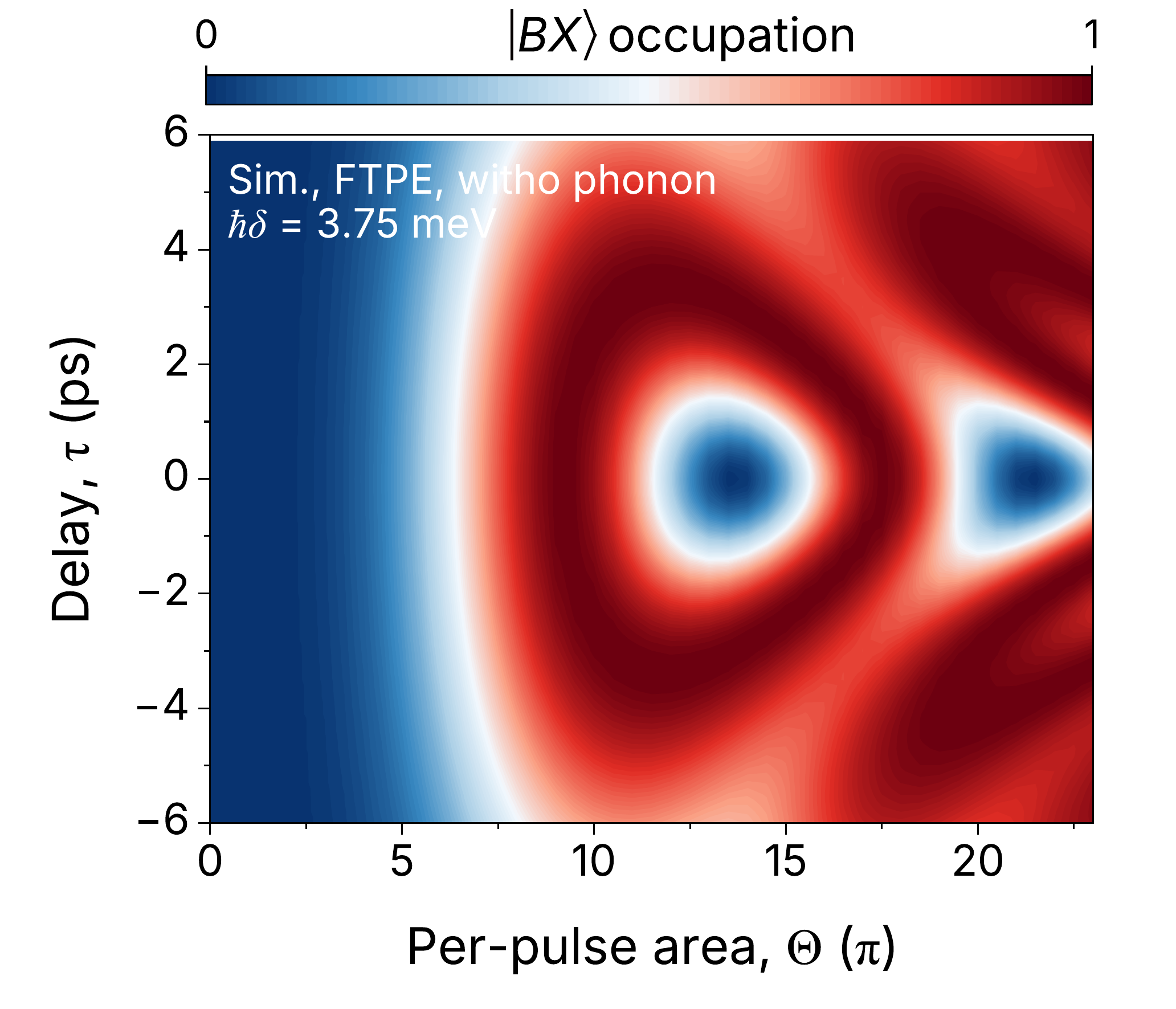}
    \caption{\textbf{Biexciton preparation efficiency at different pulse delays} Calculated $BX$ occupation versus $\Theta$ and $\tau$ without considering phonon effect.}
    \label{Sfig:delay_dep}
\end{figure}

\begin{figure}[htbp]
    \centering
    \includegraphics[width=0.9\linewidth]{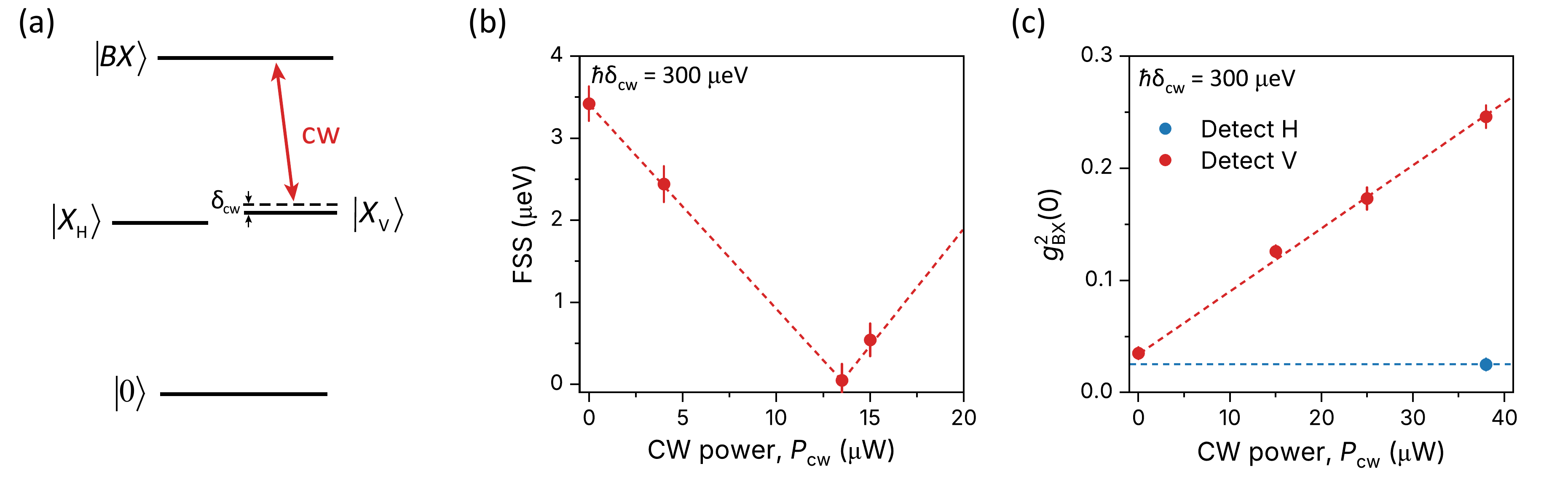}
    \caption{\textbf{FSS erasure via ac Stark effect.} \textbf{(a)}, ac Stark effect and QD energy level diagram. \textbf{(b)}, FSS as a function of applied CW laser power. Error bars arise from the standard error of sine fit residuals. \textbf{(c)}, Polarization-resolved autocorrelation measurements of $\left|BX\right>\rightarrow\left|X\right>$ photons.}
    \label{Sfig:cw fss}
\end{figure}

\section{Tuning QD FSS via ac Stark effect}
\label{Ssec:cw fss}
To eliminate the FSS of the QD, we apply a vertical linear polarized CW laser slightly detuned ($\hbar\delta_{\rm{cw}}=300~{\mu}eV$) from the $\left|BX\right>\leftrightarrow\left|X_{\rm{V}}\right>$ transition (Supplementary Fig.~\ref{Sfig:cw fss}a). This allows us to tune the FSS by varying the power of the CW laser according to:
\begin{equation}
\Delta_{\rm{FSS}}=\frac{\hbar}{2}\left(\delta_{\rm{CW}}-\sqrt{{\delta_{\rm{CW}}}^2+\Omega^2}\right).
\end{equation}

The QD investigated in the main text is QD A. Supplementary Fig.~\ref{Sfig:cw fss}(b) presents the experimentally measured FSS as a function of CW laser power. An optimal operating power of 13.5~$\mu$W is identified, where the FSS is almost eliminated. However, we observe a reduction in the single-photon purity of the $\left|BX\right>\rightarrow\left|X_{\rm{V}}\right>$ photon pair with increasing CW power (Supplementary Fig.~\ref{Sfig:cw fss}(c)). We attribute this effect to CW-induced state dressing between $\left|X_{\rm{V}}\right>$ and $\left|BX\right>$, which can lead to twice $\left|BX\right>$ emissions within a single excitation cycle. Since the CW laser exclusively couples to the V-polarized transition, the single-photon purity of the $\left|BX\right>\rightarrow\left|X_{\rm{H}}\right>$ photon remains unaffected. This limit can be overcome by further increasing the CW-laser detuning, selecting QDs with a smaller intrinsic FSS, or using other FSS compensation techniques.

\section{Calibration of pulse area, occupation, and electronic wavefunction radius}
\label{Ssec:Calibration}

To quantitatively compare the experimentally measured emission intensity with the simulated occupation, we calibrated the experimental TPE data by establishing the correspondence between the experimental observables and the theoretical parameters. As shown in Supplementary Fig.~\ref{Sfig:TPE rabi fit}, the measured TPE emission intensity was fitted to the simulated occupation as a function of pulse area. In this fitting procedure, the conversion from measured photon counts to occupation and the conversion from the square root of the excitation power to pulse area were treated as free scaling parameters. In addition, the radius of the electronic wavefunction ($a_e$) was also included as a fitted parameter in the simulation model.

The fit yields a conversion factor of ($0.0188~\rm{kcounts}^{-1}$) from the measured emission intensity to the occupation, a pulse-area calibration factor of ($0.1832~\pi/\rm{nW}^{0.5}$), and an electronic wavefunction radius of ($a_e = 5.5~\rm{nm}$). These fitted parameters were then used in all relevant analyses throughout this work.

This calibration procedure relies on the following assumptions. First, within the TPE reference dataset, the integrated emission intensity is proportional to the final occupation under fixed collection and detection conditions. Second, the experimentally controlled pulse amplitude is related to the pulse area used in the simulation by a single global scaling factor. Third, once the calibration is established from the TPE dataset, the same normalization can be consistently applied to the FTPE measurements, allowing direct comparison between experiment and theory on the same scale.

We note that since phonon effects are not very pronounced at the moderate excitation powers used for this fitting, the extracted effective phonon-coupling parameters may be subject to some uncertainty, which could introduce a small error in the calibration.

\begin{figure}[htbp]
    \centering
    \includegraphics[width=0.6\linewidth]{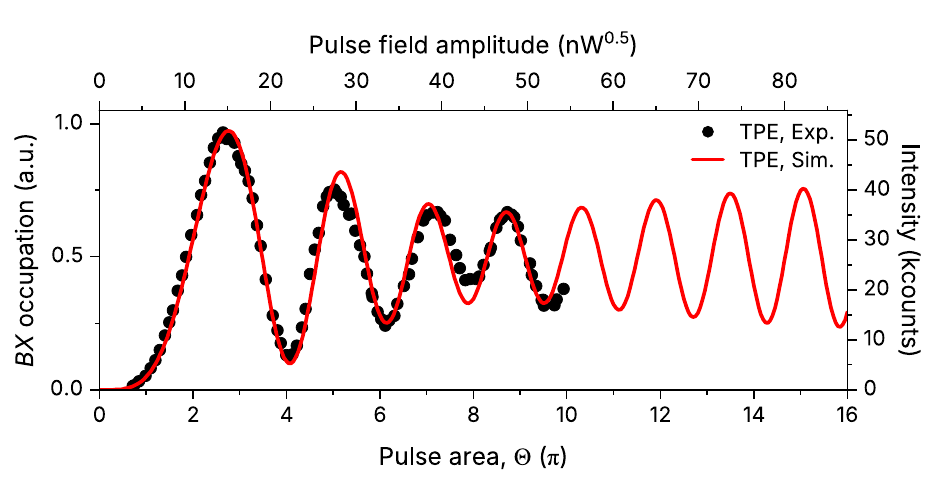}
    \caption{\textbf{Calibration of the TPE Rabi oscillation.}
    Experimental TPE Rabi oscillation data are fitted to the simulated $BX$ occupation as a function of pulse area. The fitting procedure determines the conversion factor from measured intensity to $BX$ occupation, the calibration factor from pulse field amplitude to pulse area, and the radius of the electronic wavefunction.}
    \label{Sfig:TPE rabi fit}
\end{figure}

\section{Parameters used in the calculations}
\label{Ssec:parameters}

The main parameters used in the calculations are summarized in Supplementary Table~\ref{tab:calculation_parameters}. These include the biexciton binding energy, the intrinsic fine-structure splitting, the effective electronic and hole radii, the pulse duration, the pulse detuning, and the detuning of the CW laser used for FSS compensation.

Unless otherwise specified, these parameters are fixed in the simulations presented in the main text and Supplementary Information.

\begin{table}[htbp]
\centering
\caption{Parameters used for the calculations.}
\label{tab:calculation_parameters}
\renewcommand{\arraystretch}{1.2}
\setlength{\tabcolsep}{8pt}
\begin{tabular}{lcc}
\hline\hline
Parameter & Symbol & Value \\
\hline
Biexciton binding energy & $E_b$ & 2.82~meV \\
Intrinsic fine-structure splitting & FSS & 3.4~$\upmu$eV \\
Electronic radius & $a_e$ & 5.5~nm \\
Hole radius & $a_h$ & $a_e / 1.15$ \\
Pulse electric-field duration & $\tau_0$ & 12.49~ps \\
Pulse detuning & $\hbar\delta$ & 3.75~meV \\
CW laser detuning & $\hbar\delta_{CW}$ & 300~$\upmu$eV \\
\hline\hline
\end{tabular}
\end{table}

\section{Determination of suitable detuning for FTPE in GaAs QD}
\label{Ssec:ple}

In principle, FTPE can operate over a broad range of laser detunings, as long as the two-photon resonance condition is satisfied and sufficient effective two-photon coupling is maintained. In the experiment, the practically accessible detuning range is limited by additional excitation channels in the weak-confinement GaAs/AlGaAs QDs system. To determine a suitable laser detuning for the FTPE measurements, we measured the $X$ intensity under high-power single-pulse excitation as a function of detuning. As shown in Supplementary Fig.~\ref{Sfig:PLE}, for detunings smaller than about $3$~meV, the $X$ state can still be populated through LA-phonon-assisted excitation. At larger detunings than $4.5$~meV, several resonant features appear, which we attribute to hole-state excitation~\cite{Reindl2019b}. We therefore chose a detuning of $3.75$~meV to demonstrate the reliability of the FTPE scheme as a compromise, where phonon-assisted excitation is largely suppressed while prominent hole-state excitation features are avoided.

The hole-state excitation could arise from structural asymmetry, strain-induced band mixing, and many-body configuration mixing~\cite{Narvaez2006a,Holtkemper2018}, and is not included in our four-level theoretical model. 
This restriction is specific to the present material system rather than an intrinsic limitation of the FTPE scheme. For more strongly confined InAs/GaAs QD systems, the unwanted hole-state excitation is expected to be weaker, which should allow FTPE to be implemented at larger detunings and over a broader practical detuning window.

\begin{figure}[htbp]
    \centering
    \includegraphics[width=.9\linewidth]{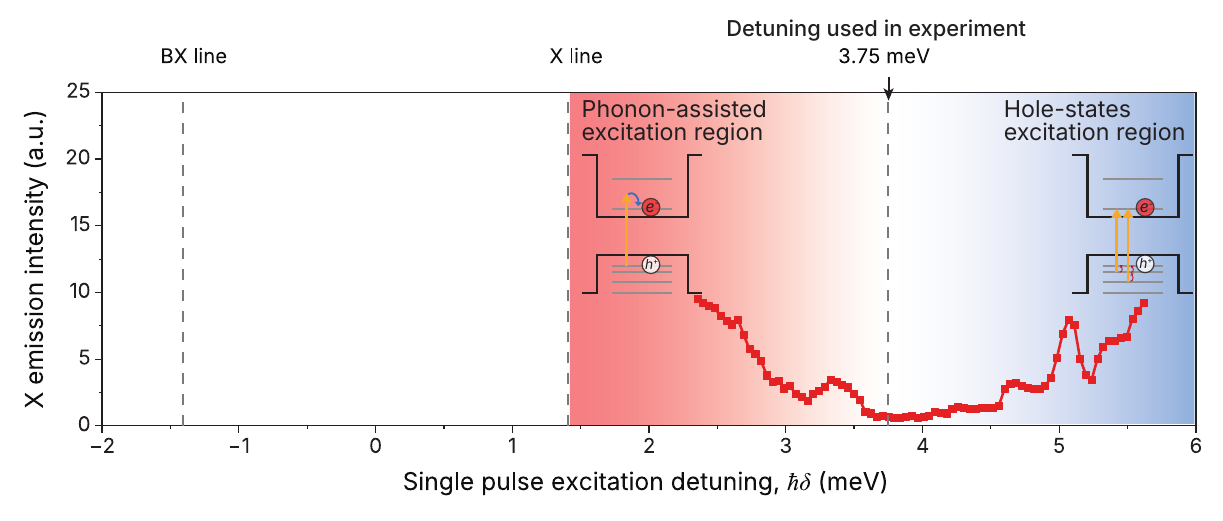}
    \caption{\textbf{High-power monochromatic pulse excitation with different detunings.}
    For detunings smaller than $\sim3$~meV, the $X$ state can be populated through LA-phonon-assisted processes, whereas at larger detunings, some population is created via hole-state excitation~\cite{Reindl2019b}.}
    \label{Sfig:PLE}
\end{figure}

\section{Validity of the Floquet model}
\begin{figure}[H]
    \centering
    \includegraphics[width=.95\linewidth]{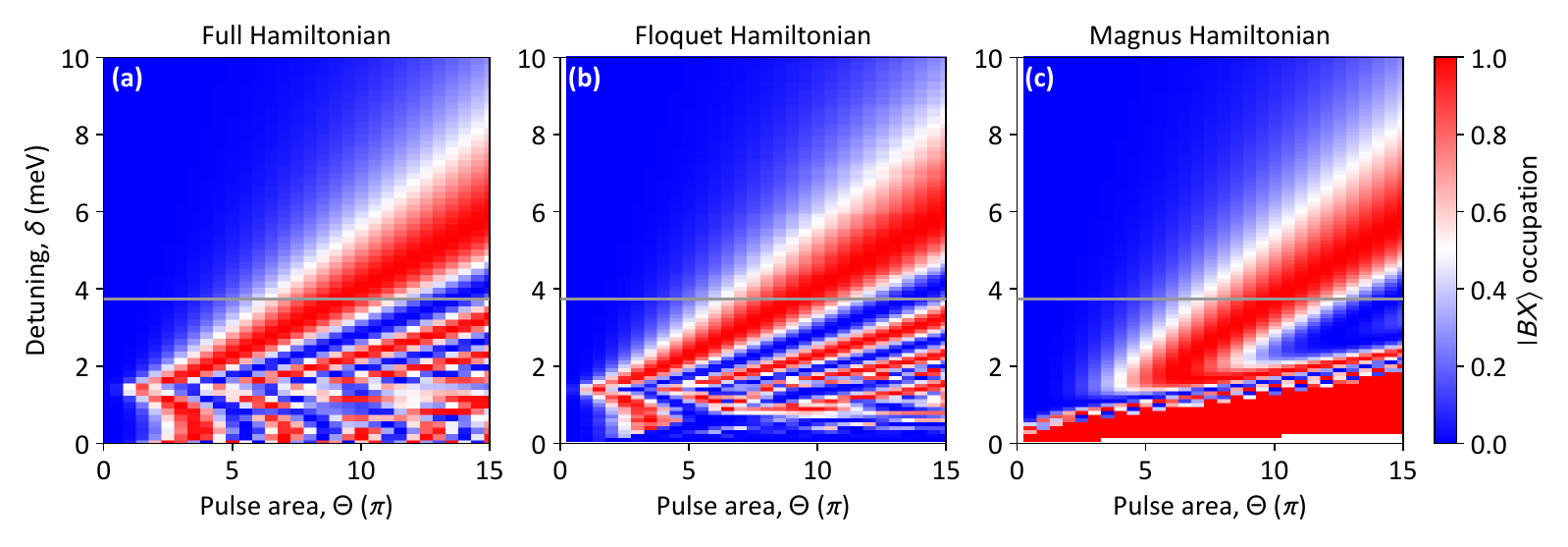}
    \caption{Pulse-area and detuning dependence of the biexciton occupation for three different levels of approximation: (\textbf{a}) the full time-dependent Hamiltonian, (\textbf{b}) the results for the numerically calculated Floquet Hamiltonian and (\textbf{c}) the results for the second-order Magnus Hamiltonian in Eq.~\eqref{eq:MagnusFloquetHamiltonian_Appendix}. All calculations are done with $\tau = 0$~ps and without phonons or losses. The gray line indicates the value of $\delta$ that is used in the experiment.}
    \label{fig:approxComp}
\end{figure}

Supplementary Fig.~\ref{fig:approxComp} shows the biexciton occupation for different levels of approximation used in our analytical investigation. In Supplementary Fig.~\ref{fig:approxComp}~a) the full time dependence is accounted for, while a numerically calculated stroboscopic Floquet Hamiltonian is used for Supplementary Fig.~\ref{fig:approxComp}~b). For Supplementary Fig.~\ref{fig:approxComp}~c), the stroboscopic Hamiltonian was obtained using second-order Magnus expansion Eq.~\eqref{eq:MagnusFloquetHamiltonian_Appendix}. For the detuning used in the experiment, which is denoted by the gray line, all three essentially match, confirming the validity of our approach.\\
Interestingly, the Floquet approximation, which assumes that the period $T = \frac{2\pi}{\delta}$ is much shorter than the time scales over which the laser envelopes change, is valid for almost the entire range of values shown in Supplementary Fig.~\ref{fig:approxComp}. In contrast, the Magnus expansion, which expands the Floquet Hamiltonian in orders of $E_b/\hbar\delta$ and/or $\Omega_{1/2}/\delta$ breaks down for $\hbar\delta \simeq E_b$ and sufficiently large pulse areas $\Theta$. The second-order expansion is sufficient to describe the first Rabi rotation, but does not reproduce higher pulse areas. Thus, if higher pulse areas or smaller binding energies are of interest, one would need to calculate the Floquet Hamiltonian numerically or use a higher order of the Magnus expansion.

\clearpage
\section{Theoretical entanglement fidelity}
\label{Ssec:simulation fidelity}
In this section, we simulate the entanglement fidelity of FTPE for $\tau = 0$~ps and $\tau = -2.9$~ps and compare the result with TPE. This is done for two cases: First, the FSS from the experiment is compensated for using a CW laser. In the second case, neither FSS nor CW laser is present. The resulting fidelities and density matrices in the polarization basis are shown in Supplementary Fig.~\ref{fig:theoryFidelity}. The simulated entanglement fidelities achieved by FTPE are close to those of TPE, though a small decrease of a fraction of a percent can be observed when using delayed pulses. Without FSS and the CW laser, the calculated fidelities are 98.2\% for TPE, 98.0\% for FTPE at $\tau = 0$~ps, and 97.7\% for FTPE at $\tau = -2.9$~ps. The slightly lower fidelity of delayed FTPE arises from the longer effective driving duration and larger pulse areas, which lead to a slightly larger pulse-induced transient spectral shift.

When the FSS is compensated for by the CW laser, the fidelity generally decreases for all considered protocols. We attribute this to the laser-induced dressing of the QD states. In particular, the exciton gains a slight biexcitonic component, which enables transitions to the second exciton state, which decreases the entanglement fidelity. This shows that the experimentally obtained value for the fidelity may be increased considerably by using other mechanisms to compensate for the FSS.

\begin{figure}[htb]
    \centering
    \includegraphics[width=\linewidth]{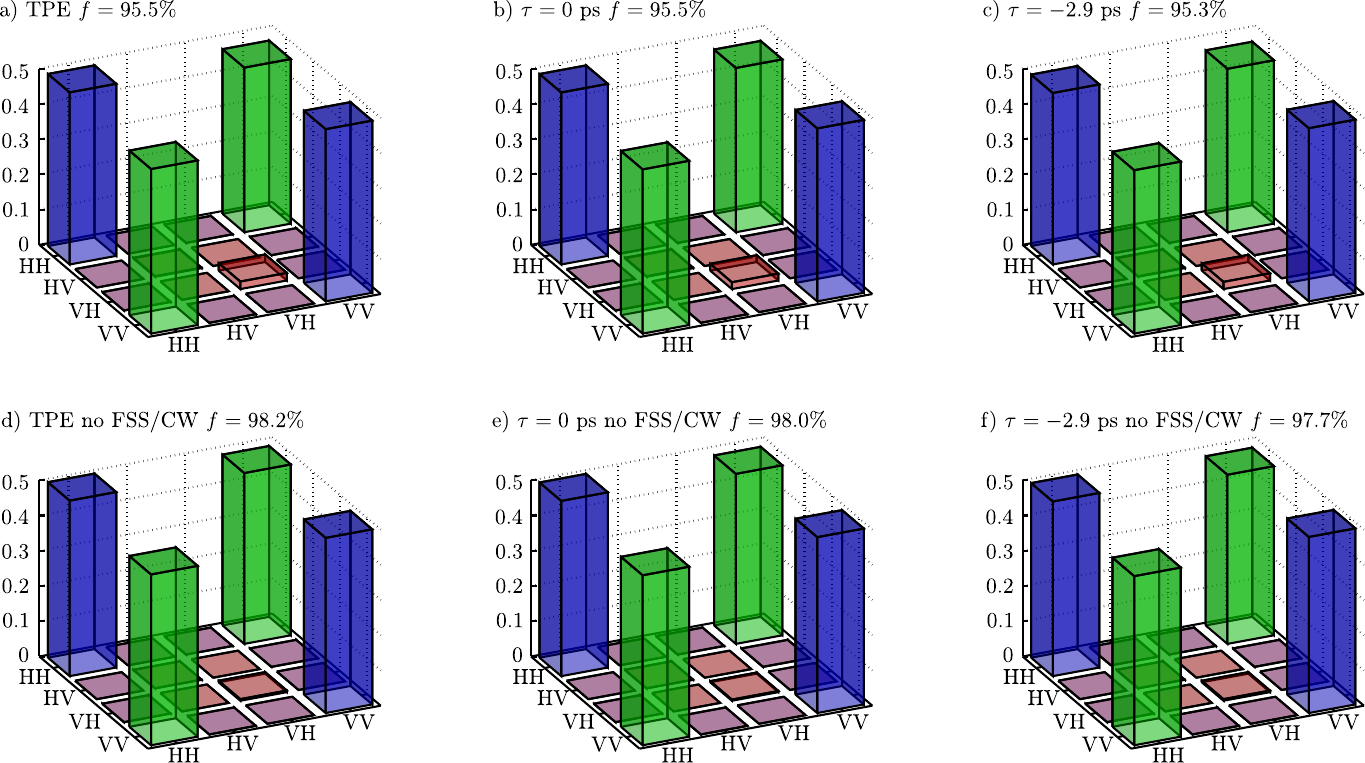}
    \caption{The density matrices and entanglement fidelities for TPE and FTPE for $\tau = 0$~ps and $-2.9$~ps with the FSS compensated by a CW laser (\textbf{a-c}) and without any FSS (\textbf{d-f}). The calculations do not account for phonons.}
    \label{fig:theoryFidelity}
\end{figure}
\newpage
\section{Robustness of FTPE against parameter variations}
\subsection{Influence of the pulse delay} \label{sec:SM_InfluenceOfDelayOnRobustness}
As shown throughout this paper, a suitable choice of the delay $\tau$ allows for robust excitation. In Supplementary Fig.~\ref{fig:differentTau_pulseAreaDetuning} we show the biexciton population as a function of pulse area and detuning for different delays. Qualitatively, we consider the excitation to be robust, when a large patch of parameters still produces high occupations. This leads to a trade-off between the demanded excitation fidelity and the size of the patch of parameters. In Supplementary Fig.~\ref{fig:differentTau_pulseAreaDetuning} we see that for our experimentally used $\delta$, which is displayed by the gray line, the excitation is robust against changes in the pulse area for delays roughly in the interval $[-2.9\text{ ps}, -3.5\text{ ps}]$.
Supplementary Fig.~\ref{fig:differentFWHM_pulseAreaDelay} displays the excitation fidelity as a function of delay time and pulse area for different pulse durations $\tau_0$. A longer pulse offers robustness at a larger pulse area.

\begin{figure}[H]
    \centering
    \includegraphics[width=\linewidth]{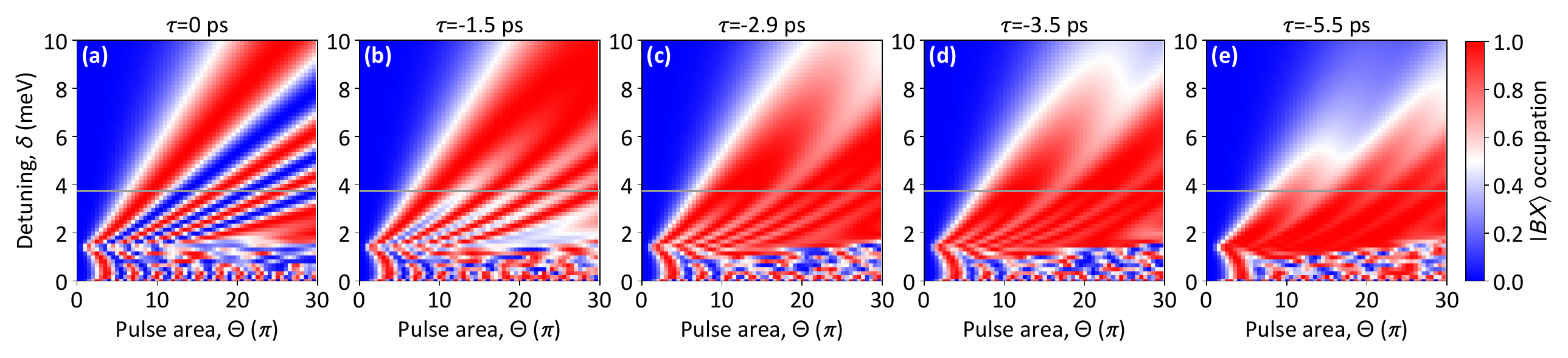}
    \caption{The pulse area-detuning heatmaps of the biexciton occupation for five different delays $\tau = 0, -1.5, -2.9, -3.5$~ps and $-5.5$~ps. The gray line indicates the detuning $\delta$ used throughout this publication. The calculations do not account for phonons.}
    \label{fig:differentTau_pulseAreaDetuning}
\end{figure}

\begin{figure}[H]
    \centering
    \includegraphics[width=\linewidth]{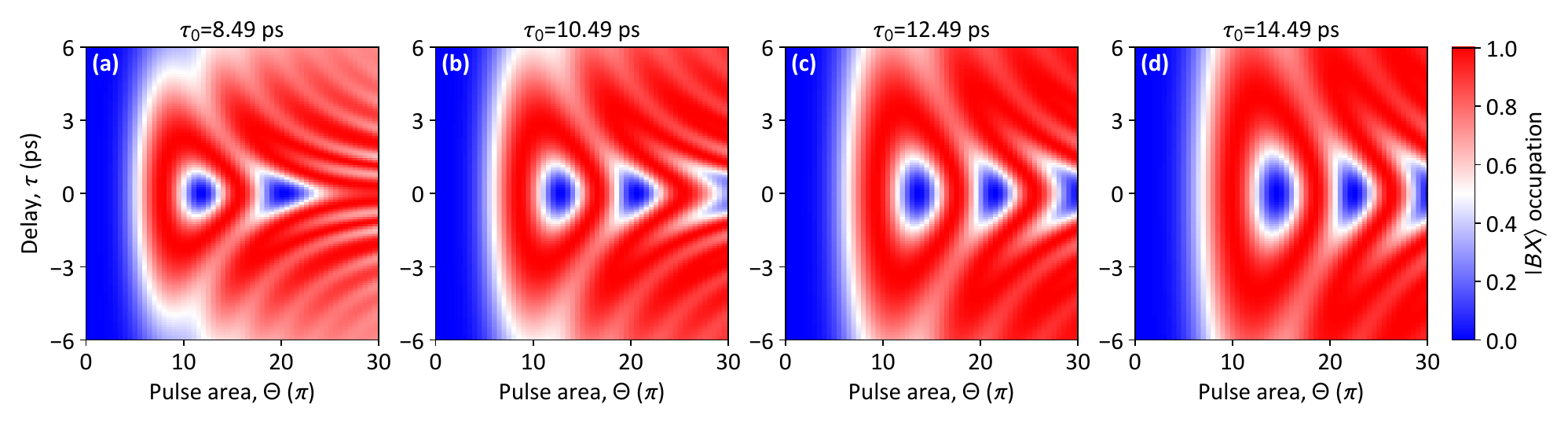}
    \caption{The pulse area-delay heatmaps of the biexciton occupation for four different pulse durations $\tau_0 = 8.49, 10.49, 12.49$~ps and $14.49$~ps. The calculations do not account for phonons.}
    \label{fig:differentFWHM_pulseAreaDelay}
\end{figure}
\newpage
\subsection{Influence of spectral shifts} \label{sec:SM_spectralShifts}
If charges near the QD shift the energy levels slightly, then the two-photon resonance that FTPE depends on can be broken. To investigate the severity of  spectral wandering, we calculate the influence of such a detuning by shifting both lasers by a frequency $\delta_e$. Biexciton population as a function of $\delta_e$ and the pulse area is shown in Supplementary Fig.~\ref{fig:bothLaserShift}. Without delay $\tau=0$, detunings of $\delta_e\simeq 0.1$~ps$^{-1}$ significantly reduce excitation efficiency. If, however, a finite delay, such as $\tau = -2.9$~ps, is used, the excitation becomes much more robust not just against changes in the pulse area, but also against deviations from the two-photon resonance.\\

\begin{figure}[H]
    \centering
    \includegraphics[width=.9\linewidth]{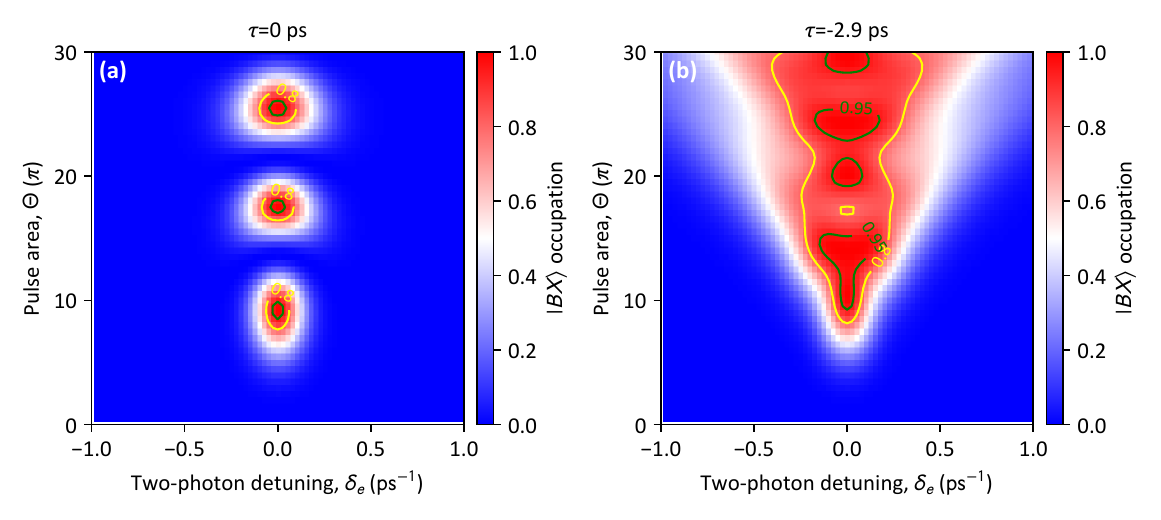}
    \caption{Biexciton occupation for different pulse areas and detunings from the 2-photon resonance for $\tau = 0$~ps and $\tau = -2.9$~ps. These calculations were done without phonons or losses.}
    \label{fig:bothLaserShift}
\end{figure}

\subsection{Influence of pulse-area imbalance}
\label{SI_Sec_intensity_imbalance}

Equal peak intensities of the red- and blue-detuned pulses are not a strict requirement for FTPE. Nevertheless, equal driving strengths constitute the natural symmetric operating condition and are typically close to optimal. An imbalance between the two pulse amplitudes breaks this symmetry and modifies both the effective coupling and the associated stroboscopic evolution. Consequently, the optimal working point and the corresponding robustness window may shift, while high-fidelity biexciton preparation can still be achieved.

\begin{figure}[H]
    \centering
    \includegraphics[width=.9\linewidth]
    {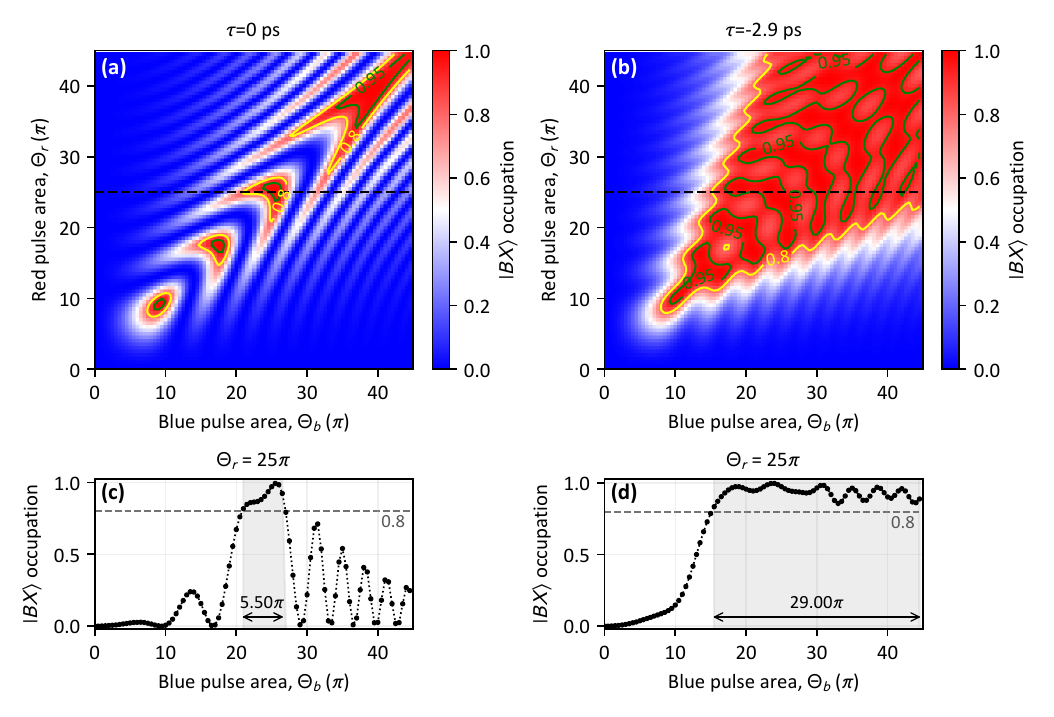}
    \caption{
    Final biexciton population after FTPE as a function of the independently varied driving strengths of the red- and blue-detuned pulses for $\tau=0$~ps (\textbf{a}) and $\tau=-2.9$~ps (\textbf{b}). The calculations were done without phonons or losses. Contour lines for the biexciton occupation $80\%$ and $95\%$ are shown using yellow and green lines, respectively. (\textbf{c,d}) Line cuts of (\textbf{a,b}), respectively, at a fixed red-pulse area of $\Theta_r=25\pi$, showing the biexciton occupation as a function of the blue-pulse area. The shaded regions indicate the ranges of blue-pulse area over which the biexciton occupation exceeds $80\%$.
    }
    \label{fig:varyBothAmplitudes}
\end{figure}

To quantify the tolerance of FTPE to such an imbalance, we independently vary the driving strengths of the red- and blue-detuned pulses. Supplementary Fig.~\ref{fig:varyBothAmplitudes} shows the resulting final biexciton population for simultaneous pulses, $\tau=0$~ps, and for a finite delay, $\tau=-2.9$~ps. The calculations are performed without phonon coupling or radiative losses.

For $\tau=0$~ps, the high-population region is relatively narrow under independent variations of the two pulse strengths. By contrast, for $\tau=-2.9$~ps, the region with high biexciton occupation extends over a substantially wider range of pulse-strength combinations. To quantify this difference more directly, Supplementary Fig.~\ref{fig:varyBothAmplitudes}(c,d) shows line cuts through the corresponding maps at a fixed red-pulse area of $\Theta_r=25\pi$. For $\tau=0$~ps, the range of blue-pulse areas over which the biexciton occupation remains above $80\%$ is only $5.5\pi$. For $\tau=-2.9$~ps, this range increases to $29\pi$.

These results show that the pulse-area robustness of FTPE is not restricted to simultaneous variations of both pulse strengths, but is also retained when the two pulse areas vary independently. Equal peak intensities should therefore be regarded as a convenient near-optimal operating condition rather than a fundamental requirement of FTPE.
\newpage

\newpage
\section{Comparison with other excitation mechanisms}
\label{diff_to_others}
\subsection{Nearly monochromatic two-photon resonant excitation}\label{sec:SM_tpe}
While conventional monochromatic two-photon excitation is described by the same Hamiltonian as in Eq. (1) in the limit of zero detuning $\delta\to 0$, it is important to note that FTPE and TPE are qualitatively different and appear in opposite parameter regimes. It is instructive to consider the prediction of the usual description of TPE~\cite{Stufler2006} when a finite detuning $\delta$ is accounted for. Here, we focus on the case of zero delay $\tau=0$ between the pulses, where the driving is given by 
\begin{equation}
\Omega(t)=2f(t)\cos(\delta t).
\end{equation}

In contrast to the description of FTPE valid for large detunings, TPE is described using an adiabatic approximation~\cite{Stufler2006}. There the initial state $|G\rangle$ is decomposed into laser-dressed eigenstates. Over the pulse duration, the occupation of the instantaneous eigenstates is assumed to be constant by the adiabatic theorem, while the respective components of the wave function acquire dynamical phases corresponding to the time integral over the respective energy eigenvalues. The final biexciton state occupation $|\langle BX|\Psi\rangle|^2$ is then the result of the interference of the different eigenstate components of the wave function. One finds~\cite{Stufler2006}

\begin{equation} \label{eq:adiabaticTPEFormula}
    |\langle BX|\psi\rangle|^2 = \text{sin}^2\,\frac{\Lambda}{2},
\end{equation}
with
\begin{equation}
    \Lambda = \frac{1}{4\hbar}\int_{-\infty}^{\infty}\left[E_b-\sqrt{E_b^2+4\cdot 8\hbar^2\Omega^2(t)}\right]\,\text{d}t.
\end{equation}

The final biexciton state occupations are depicted in Supplementary Fig.~\ref{fig:compToTPE} for both direct numerical simulations of the dynamics and the results of the adiabatic approximation. Indeed, for small detunings $\delta\ll E_b/(2\hbar)$, the adiabatic approximation captures the Rabi-like behaviour of the full dynamics well. However, for larger detunings, here $\delta\gtrsim E_b/(4\hbar)$, the adiabatic approximation starts to break down and the distribution of parameter sets for which a significant biexciton population is achieved becomes erratic.

Clearly, at $\delta=E_b/(2\hbar)$ the lasers are resonant with direct transitions involving the single exciton state. Thus, even though near TPE and FTPE have in common that they involve virtual transitions without significant occupation of the exciton state, they are clearly distinguished by the parameter regimes $\delta < E_b/(2\hbar)$ and $\delta > E_b/(2\hbar)$. Thus, the physics in both regimes is different,
which is why a different theoretical approach is required to explain FTPE.\\

\begin{figure}[htbp]
    \centering
    \includegraphics[width=.85\linewidth]{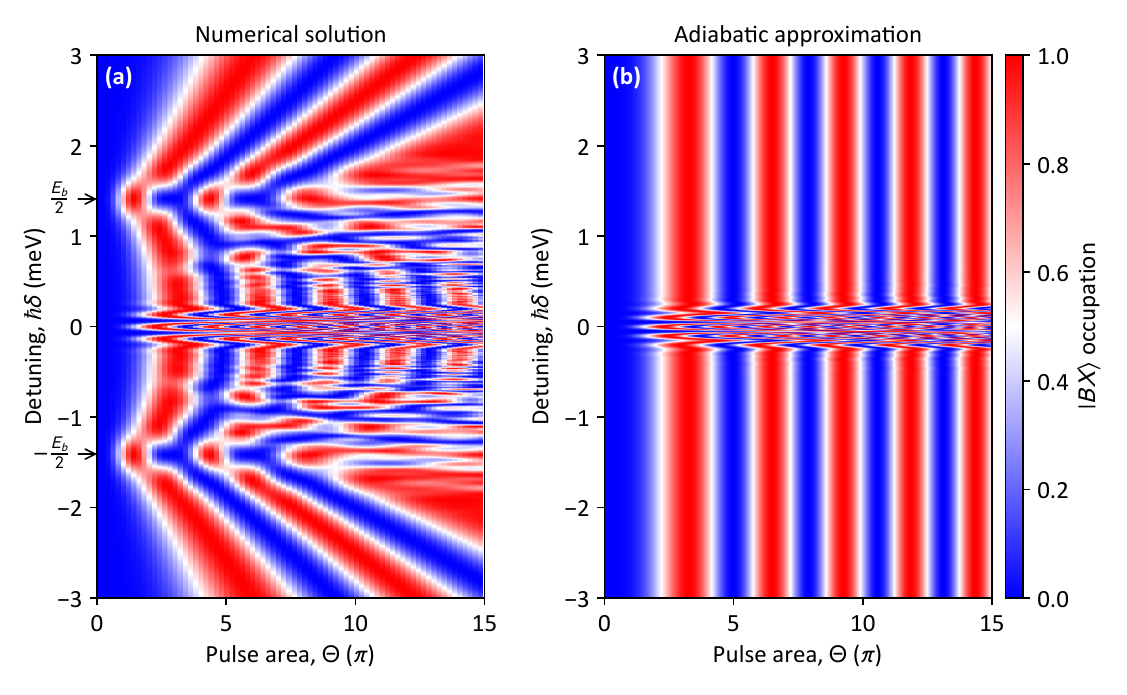}
    \caption{Numerically calculated final biexciton population under near-two-photon resonant excitation as a function of detuning $\delta$ and pulse area per pulse $\Theta$ obtained by (\textbf{a}) direct numerical simulation of the three-level dynamics without phonons or losses and (\textbf{b}) using the adiabatic approximation Eq.~\eqref{eq:adiabaticTPEFormula}. Half the biexciton binding energy is chosen as $E_b/2 = 1.41\text{ meV}$ and quantum dot-phonon interactions are neglected in both calculations.}
    \label{fig:compToTPE}
\end{figure}

 % The approximation only fits well for small detunings $\delta \ll E_b/2\hbar$

\subsection{Adiabatic protocols}
A number of state preparation protocols for quantum dots have been proposed that rely on adiabatic driving, such as adiabatic rapid passage using a chirped laser pulse~\cite{Gawarecki2013}, whose frequency changes over time, or ultrafast electric tuning~\cite{Mukherjee2020}. Also phonon-assisted state preparation employs adiabatic undressing~\cite{Barth2016b} as a working principle. 
In particular, adiabatic protocols with the state of the systems confined to the lowest instantaneous eigenstate for most of the evolution are robust against small variations, e.g., of the pulse area. This raises the question of whether the robustness found for FTPE is related to adiabatic transitions. 

\begin{figure}[htbp]
    \centering
    \includegraphics[scale=0.9]{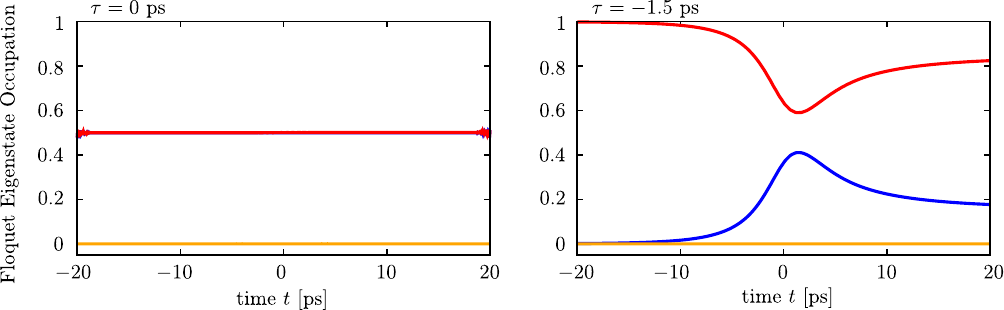}
    \caption{\label{fig:notadiabatic}Time evolution of the occupation of all three eigenstates of the stroboscopic Hamiltonian $\bar{H}$. Displayed are cases without any delay, $\tau = 0$~ps, as well as a small negative delay, $\tau=-1.5$~ps. Adiabatic transitions require occupations of instantaneous eigenstates to be constant in time, which is not fulfilled for finite delays.}
\end{figure}

To this end, we depict in Supplementary Fig.~\ref{fig:notadiabatic} the occupations of instantaneous eigenstates of the stroboscopic Hamiltonian for parameters $\hbar\delta=3.75$~meV and $E_b/2=1.41$~meV for cases without ($\tau=0$~ps) and with ($\tau=-1.5$~ps) a delay between the pulses for FTPE. For finite delay, it can be seen that the dynamics is clearly nonadiabatic in the instantaneous eigenbasis of the stroboscopic Hamiltonian. Nevertheless, the dynamics is found to be robust. For vanishing delay, the occupations of instantaneous eigenstates remain nearly constant, similar to the behaviour expected in the adiabatic limit. But because the system is in a coherent superposition between two eigenstates, the relative phase is generally sensitive to variations in the pulse area, making the protocol not more robust than, e.g., Rabi driving. To summarize, in the Floquet regime, the dynamics may or may not be adiabatic, depending on the pulse delay, and the enhanced robustness is mainly found in the diabatic regime. This shows that the physics of FTPE differs from that of conventional adiabatic protocols.

\subsection{STIRAP}
A state preparation scheme whose setup, on first glance, resembles FTPE is stimulated Raman adiabatic passage (STIRAP)~\cite{Bergmann2015}. STIRAP is designed to drive transitions between two ground states via a third, intermediate state. One assumes selection rules where the first (pump) pulse couples only the first ground state to the intermediate state and the second (Stokes) pulse couples the intermediate state to the second ground state.
Furthermore, both lasers are detuned from the respective transition with the same detuning $\Delta$. In the basis of the second ground, intermediate, and first ground states, the corresponding Hamiltonian is given by
\begin{equation}
    H^{\text{STIRAP}} = \hbar\left(\begin{array}{ccc}
       0  & \frac{\Omega_p}{2} & 0 \\
       \frac{\Omega_p}{2}  & \Delta & \frac{\Omega_s}{2}\\
       0 & \frac{\Omega_s}{2} & 0\\
    \end{array}\right).
\end{equation}
where $\Omega_p(t)$ and $\Omega_s(t)$ are the (real) amplitudes of the pump and Stokes pulses, respectively. 
The Hamiltonian is similar to Eq.~(1) when the detunings are identified by $\Delta=\delta-E_b/(2\hbar)$. 
The delay and pulse area dependence of both STIRAP and FTPE for these parameters are depicted in Supplementary Fig.~\ref{fig:compToSTIRAP}. 
Both show Rabi rotations for zero delay and broad parameter regimes with large final state occupations.

\begin{figure}[htbp]
    \centering
    \includegraphics[width=.8\linewidth]{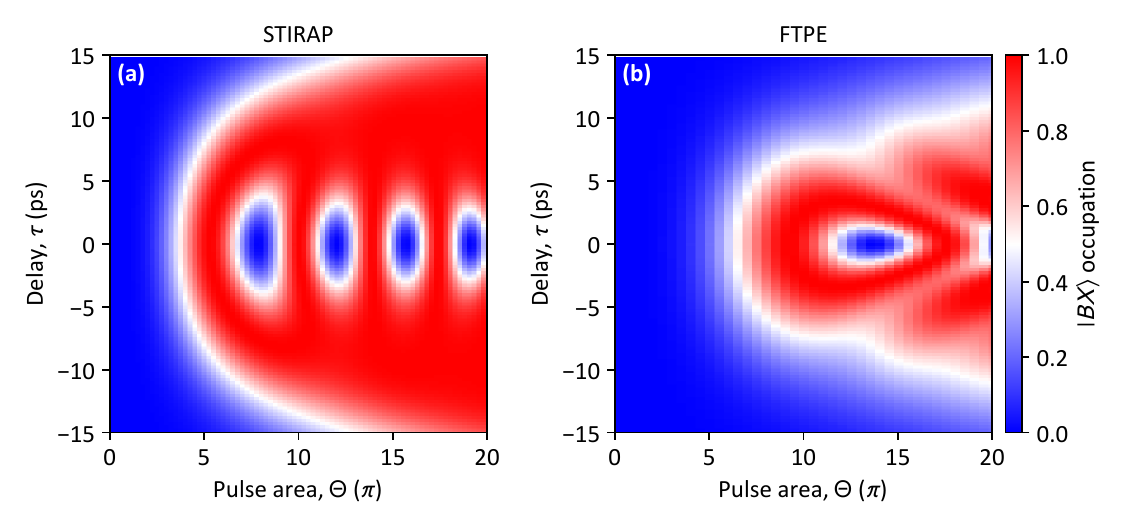}
    \caption{Comparison of the $\tau$-$\Theta$ dependence of the biexciton occupation for regular STIRAP and FTPE. $\Theta$ is the pulse area of either individual laser. The detuning to the central level was chosen $\hbar\Delta = 2.34\text{ meV}$ in both cases. The calculations were done without phonons or losses.}
    \label{fig:compToSTIRAP}
\end{figure}

However, there are also clear differences between STIRAP and FTPE. Most importantly, for STIRAP, the transitions can be controlled separately by the two laser pulses. The optical selection rules in the biexciton-exciton system of a quantum dot, on the other hand, are such that linearly polarized lasers couple both transitions simultaneously by both pulses. Even in the case of two lasers with oppositely circularly polarized pulses, for which the first pulse couples the ground state to an intermediate exciton state in the circular polarization basis and the second pulse couples the same exciton state to the biexciton state, the situation deviates from STIRAP, since for finite overlap of the pulses transitions to the exciton state with opposite polarization (a fourth level) are possible. Either way, the optical selection rules for semiconductor quantum dots prohibit the straightforward implementation of STIRAP for biexciton state preparation.

\subsection{SUPER and dichromatic excitation}
Recently, two-colour excitation schemes have been explored for state preparation of quantum dots, in particular, exciton state preparation for applications in on-demand single-photon generation~\cite{He2019,Koong2021}. An important goal of off-resonant excitation is that photons from the excitation laser can be easily spectrally separated from the emitted quantum state of light. A detailed analysis of the Swing-UP of quantum Emitter Population (SUPER) scheme has revealed that parameters for which off-resonant two-colour excitation leads to near-unity population inversion are found when one laser becomes resonant with the transition between the laser-dressed states Stark-shifted by the other laser~\cite{Bracht2021}.

FTPE has in common with SUPER mainly the fact that two-colour off-resonant excitation is employed. But in contrast to SUPER, Stark shifts do not set the resonance condition or constitute the working mechanism of FTPE. For the same reason, FTPE also differs from dichromatic excitation with one red- and one blue-detuned laser~\cite{Koong2021}.\\
In Ref.~\cite{Bracht2023a}, the use of SUPER for the excitation of the biexciton state was investigated. Starting with the same parameters as in Ref.~\cite{Bracht2023a}, we vary both pulse areas and depict the resulting biexciton population in Supplementary~Fig.~\ref{fig:SUPER_Heatmap} for three different pulse delays. Compared to the corresponding results for FTPE, which are displayed in Supplementary~Fig.~\ref{fig:varyBothAmplitudes}(b), SUPER is more sensitive to variations in the resonance conditions, as can be seen by the fact that the area enclosed in the 80\% and the 95\% is considerably larger for FTPE.

\begin{figure}
    \centering
    \includegraphics[width=1\linewidth]{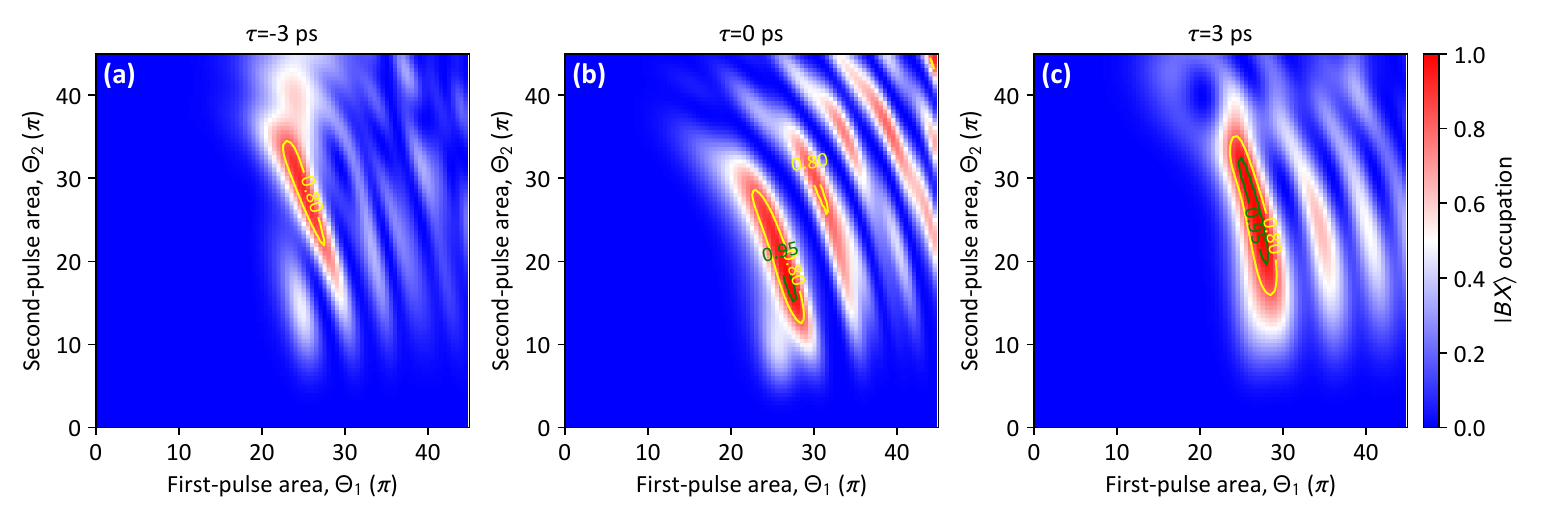}
    \caption{Biexciton occupation of SUPER excitation using parameters from Ref.~\cite{Bracht2023a} for different pulse areas and delays. The calculations were done without phonons or losses.}
    \label{fig:SUPER_Heatmap}
\end{figure}

\newpage
\section{Relative phase difference between the lasers} 
\label{SI_Sec_phserelation}

It is possible to add a relative phase $\phi$ to one of the lasers. Then, the laser function reads
\begin{equation} \label{eq:laserWithRelPhase}
    \Omega(t) = f(t-\tau/2)\text{e}^{-\text{i}\delta (t-\tau/2)} + f(t+\tau/2)\text{e}^{\text{i}(\delta (t+\tau/2)+\phi)}.
\end{equation}
The numerical investigations in Supplementary Fig.~\ref{fig:relPhase}(a) show that the relative phase between the lasers is irrelevant for the final occupation. Our protocol produces the same excitation regardless of the specific value of $\phi$. It is similar to SUPER in this aspect. But, again like for SUPER, the oscillations during the excitation are affected, as can be seen in Supplementary Fig.~\ref{fig:relPhase}(b).

\begin{figure}[htbp]
    \centering
    \includegraphics[width=.8\linewidth]{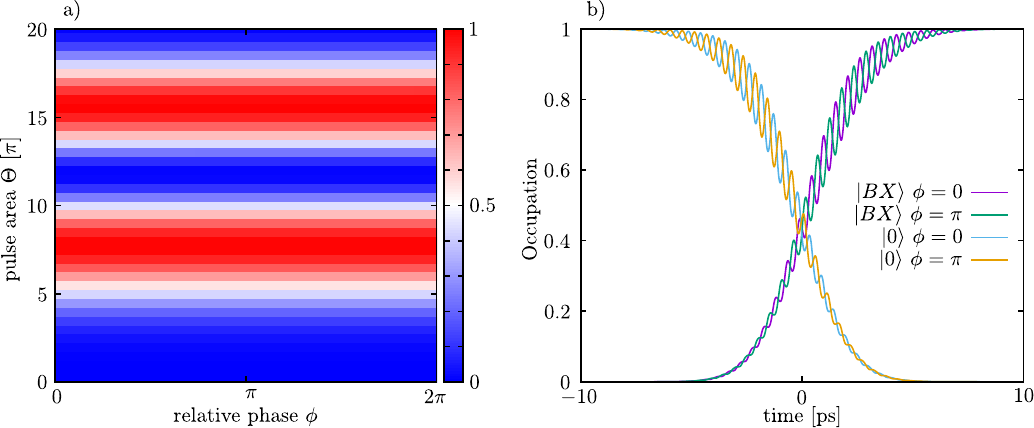}
    \caption{\textbf{(a)}, Biexciton occupation after using two pulses with pulse area $\Theta$ in which one has the relative phase $\phi$, as in Eq.~\eqref{eq:laserWithRelPhase}. $\hbar\delta = 3.75\text{ meV}$ was chosen as a typical value and no delay between the lasers was used. The occupation clearly does not depend on the relative phase between the pulses. We find that the same applies in cases of $\tau \neq 0$ (not shown). \textbf{(b)}, The occupation of ground and biexciton states for two exemplary phases $\phi \in \{0,\pi\}$.}
    \label{fig:relPhase}
\end{figure}

\end{bibunit}
\end{document}